\documentclass{aa}
\usepackage{graphicx}
\usepackage{txfonts}
\usepackage{amsmath,amssymb}
\usepackage{mathrsfs}   
\usepackage{mathtools}  
\usepackage{color}
\usepackage{colortbl}
\definecolor{Gray}{gray}{0.9}
\newcolumntype{a}{>{\columncolor{Gray}}l}
\newcolumntype{b}{>{\columncolor{Gray}}r}
\usepackage{array}

\usepackage{natbib}
\bibpunct{(}{)}{;}{a}{}{,} 

\newcommand{\fj}{f_j}
\newcommand{\fM}{f_M}
\newcommand{\fv}{f_V}
\newcommand{\fr}{f_R}
\newcommand{\fb}{f_{\rm b}}
\newcommand{\fzero}{f_0}
\newcommand{\vzero}{V_0}
\newcommand{\mzero}{M_0}
\newcommand{\lfzero}{\log\,f_0}
\newcommand{\lvfzero}{\log\,V_0}
\newcommand{\vf}{V_{\rm flat}}
\newcommand{\Mh}{M_{\rm h}}
\newcommand{\Vh}{V_{\rm h}}
\newcommand{\Rh}{R_{\rm h}}
\newcommand{\jh}{j_{\rm h}}

\newcommand{\rd}{R_{\rm d}}
\newcommand{\Msun}{M_\odot}
\newcommand{\mstar}{M_\star}
\newcommand{\Mstar}{M_\star}
\newcommand{\mbary}{M_{\rm baryons}}
\newcommand{\mhi}{M_{\rm HI}}
\newcommand{\jstar}{j_\star}

\newcommand{\Omegab}{\Omega_{\rm b}}

\newcommand{\Omegam}{\Omega_{\rm m}}
\newcommand{\jbar}{j_{\rm baryon}}
\newcommand{\siglf}{\sigma_{\log\,f}}
\newcommand{\siglm}{\sigma_{\log\,M_\star}}
\newcommand{\siglr}{\sigma_{\log\,R_{\rm d}}}
\newcommand{\siglj}{\sigma_{\log\,j_\star}}
\newcommand{\siglam}{\sigma_{\log\,\lambda}}
\newcommand{\de}{{\rm d}}

\defcitealias{SPARC}{LMS16}
\defcitealias{PFM19}{PFM19}
\defcitealias{Moster+13}{M+13}

\begin{document}

\title{Galaxy disc scaling relations: A tight linear galaxy--halo connection challenges abundance matching}
\titlerunning{Galaxy disc scaling relations}
\authorrunning{L. Posti, A. Marasco, F. Fraternali \& B. Famaey}

  \author{Lorenzo Posti\inst{1}\fnmsep\thanks{lorenzo.posti@astro.unistra.fr},
          Antonino Marasco\inst{2,3}
          \and
          Filippo Fraternali\inst{2}
          \and
          Benoit Famaey\inst{1}
          }
  \institute{Universit\'e de Strasbourg, CNRS UMR 7550, Observatoire astronomique de Strasbourg, 11 rue de l'Universit\'e, 67000 Strasbourg, France.
         \and
             Kapteyn Astronomical Institute, University of Groningen,
                  P.O. Box 800, 9700 AV Groningen, the Netherlands
         \and
             ASTRON, Netherlands Institute for Radio Astronomy, Oude Hoogeveensedijk 4, 7991 PD, Dwingeloo, The Netherlands
             }
      \date{Received XXX; accepted YYY}

  \abstract{In $\Lambda$CDM cosmology, to first order, galaxies form out of the cooling of baryons within the virial radius of their dark matter halo. The fractions of mass and angular momentum retained in the baryonic and stellar components of disc galaxies put strong constraints on our understanding of galaxy formation. In this work, we derive the fraction of angular momentum retained in the stellar component of spirals, $\fj$, the global star formation efficiency $\fM$, and the ratio of the asymptotic circular velocity ($\vf$) to the virial velocity $\fv$, and their scatter, by fitting simultaneously the observed stellar mass-velocity (Tully-Fisher), size-mass, and mass-angular momentum (Fall) relations. We compare the goodness of fit of three models: (i) where the logarithm of $\fj$, $\fM$, and $\fv$ vary linearly with the logarithm of the observable $\vf$; (ii) where these values vary as a double power law; and (iii) where these values also vary as a double power law but with a prior imposed on $\fM$ such that it follows the expectations from widely used abundance matching models. We conclude that the scatter in these fractions is particularly small ($\sim 0.07$~dex) and that the “linear” model is by far statistically preferred to that with abundance matching priors. This indicates that the fundamental galaxy formation parameters are small-scatter single-slope monotonic functions of mass, instead  of  being  complicated  non-monotonic  functions. This incidentally confirms that the most massive spiral galaxies should have turned nearly all the baryons associated with their haloes into stars. We call this the failed feedback problem.
             }
  \keywords{galaxies: kinematics and dynamics -- galaxies: spiral -- galaxies: structure --
                         galaxies: formation}
  \maketitle

\section{Introduction} \label{sec:intro}

The current $\Lambda$ cold dark matter ($\Lambda$CDM) cosmological model is very successful
at reproducing observations of the large-scale structure of the Universe. However, galactic
scales still present to this day a number of interesting challenges for our understanding of
structure formation in such a cosmological context \citep[e.g.][]{Bullock}. These challenges
could have important consequences on our understanding of the interplay between baryons and
dark matter, or even on the roots of the cosmological model itself, including the very nature
of dark matter. For instance, the most inner parts of galaxy rotation curves present a wide
variety of shapes \citep{Oman1,Oman2}, which might be indicative of a variety of central
dark matter profiles ranging from cusps to cores and closely related to the observed central
surface density of baryons \citep[e.g.][]{LelliFrat,LellicentralDM,Ghari}. In addition to
such surprising central correlations, the phenomenology of global galactic scaling laws,
which relate fundamental galactic structural parameters of both baryons and dark matter, also
carries important clues that should inform us about the galaxy formation process in a
cosmological context.

Given the complexity of the baryon physics leading to the formation of galaxies, which
involves for instance gravitational instabilities, gas dissipation, mergers
and interactions with neighbours, or feedback from strong radiative sources, it is
remarkable that many of the most basic structural scaling relations of disc galaxies are
simple, tight power laws \citep[see e.g.][for a review]{vanderKruitFreeman11}; these most
basic structural scaling relations, for example, can be
between the stellar or baryonic mass of the galaxy and its rotational velocity
\citep[][]{TullyFisher77,Lelli+16a}, its stellar mass and size \citep[][]{Kormendy77,Lange+16},
and its stellar mass and stellar specific angular momentum \citep[][]{Fall83,Posti+18b}.

The interplay of all the complex phenomena involved in the galaxy formation process thus
conspires to produce a population of galaxies which is, to first order, simply rescalable.
Interestingly, in $\Lambda$CDM, dark matter haloes also follow simple, tight, power-law
scaling relations and their structure is fully rescalable. Thus, all of this is
suggestive of the existence of a simple correspondence between the scaling relations of dark
matter haloes and galaxies \citep[e.g.][]{Posti+14}. In this context, we can consider to
first order a simplified picture in which galaxies form out of the cooling of baryons within
the virial radius of their dark matter halo. That is, before
any dissipation happens, the fraction of total matter that is baryonic inside newly
formed haloes would not differ on average from the current value of the cosmic baryon
fraction $\fb\equiv\Omegab/\Omegam \simeq 0.157$, where $\Omegab$ and $\Omegam$ are
the baryonic and total matter densities of the Universe, respectively \citep[][]{Planck18}.
In this simplified picture, galaxies are then formed out of those baryons that effectively
dissipate and sink towards the centre of the potential well, and the final structural
properties of galaxies, such as mass, size, and angular momentum, are then directly
related to the interplay between the (cooling) baryons and (dissipationless) dark matter.

Observationally proving that indeed the masses, sizes, and angular momenta of galaxies are
simply and directly proportional to those of their dark matter haloes, would be a major
finding. This means that, out of all the complexity of galaxy formation in a cosmological
context, a fundamental regularity is still emerging, which we would then need to understand.
Some of the earliest and most influential theoretical models of disc galaxy
formation relied on reproducing the observed scaling laws of discs to constrain their
free parameters \citep[e.g.][]{FallEfstathiou80,Dalcanton+97,MMW98}. These parameters
are often chosen to be physically meaningful and fundamental quantities that
synthetically encode galaxy formation, such as a global measure of the efficiency
at forming stars from the cooling material \citep[e.g.][hereafter
\citetalias{Moster+13}]{Behroozi+13,Moster+13} or a measure of the net gains or losses
of the total angular momentum from that initially acquired via tidal torques
\citep[][]{Peebles69,RF12,Pezzulli+17}.
The rich amount of data collected in recent years for spiral galaxies both in the
nearby Universe and at high redshift allows an unprecedented exploitation of the
observed scaling laws which, when fitted simultaneously, can yield very strong
constraints on such fundamental galaxy formation parameters \citep[e.g.][]{Dutton+07,
DuttonvdBosch12,DesmondWechsler15,Lapi+18}.

While being the focus of many studies over the past years, the connection between
galaxy and halo properties is still not trivial to measure observationally
\citep[see][for a recent review]{WechslerTinker18}. However, arguably the most
important bit of this connection, the relation between galaxy stellar mass and
dark matter halo mass, is very well studied and the results from different groups
tend to converge towards a complex, non-linear correspondence. As long as
galaxies of all types are considered and stacked together, the same non-linear
relation, with a break at around $L^\ast$ galaxies, is found irrespective of
the different observations used to probe this relation:\ for instance, the match of the halo
mass function to the observed stellar mass function (the so-called
abundance matching ansatz; e.g. \citealt{ValeOstriker04,Kravtsov+04,Behroozi+13};
\citetalias{Moster+13}), galaxy clustering \citep[e.g.][]{Zheng+07}, group catalogues
\citep[e.g.][]{Yang+08}, weak galaxy-galaxy lensing \citep[e.g.][]{Mandelbaum+06,
Leauthaud+12,vanUitert+16}, and satellite kinematics \citep{vandenBosch+04,More+11,
WojtakMamon13}. Hence, this would imply that the high regularity of the observed disc
scaling laws is not a direct reflection of the rescalability of dark matter
haloes. If the stellar-to-halo mass relation of disc galaxies is non-linear,
then the relation between disc rotation velocity and halo virial velocity
\citep{NavarroSteinmetz00,Cattaneo+14,Ferrero+17}, as well as the
relation between stellar and halo specific angular momenta \citep{Shi+17,Posti+18a},
also have to be highly non-linear.

Nonetheless, recently \citet[][hereafter \citetalias{PFM19}]{PFM19} measured
individual halo masses for a large sample of nearby disc galaxies, from small
dwarfs to spirals $\sim 10$ times more massive than the Milky Way. These authors used
accurate near-infrared (3.6 $\mu$m) photometry with the Spitzer Space Telescope
and HI interferometry
\citep{SPARC} to determine the stellar and dark matter halo masses robustly,
by means of fitting the observed gas rotation curves. Surprisingly, the authors found
 no indication of a break in the stellar-to-halo mass relation of their
sample of spirals. This finding is in significant tension with expectations of abundance
matching models for galaxies with stellar masses above $8\times
10^{10}\Msun$ \citep[see also][]{McGaugh+10}. Since the high-mass slope
of the stellar-to-halo mass relation is commonly understood in terms of strong
central feedback \citep[e.g.][]{WechslerTinker18}, we call this observational
discrepancy the failed feedback problem.
This discrepancy might be there simply as a result of a
morphology-dependent galaxy-halo connection. While the relation
found by \citetalias{PFM19} applies to disc galaxies, the stellar-to-halo
mass relation from abundance matching instead is an average statistic derived
for galaxies of all types that is heavily dominated by spheroids at the
high-mass end. This would imply that the galaxy-halo connection for discs
and spheroids can be significantly different, for example it could be linear for
discs while being highly non-linear for spheroids.

If this is the case for disc galaxies in the nearby Universe,
then this should leave a measurable imprint on their structural scaling
laws, such as the Tully-Fisher, size-mass, and Fall\footnote{
We call the relation between stellar mass and stellar specific angular
momentum the ``Fall relation'' hereafter, due to the pioneering work
by \cite{Fall83} who realised the importance of this law in galaxy
formation.
} relations. It is possible to model these three scaling laws
(of which the last two are dependent) with three (dependent) fundamental
galaxy formation parameters: one to determine the stellar-to-halo mass
relation, one for the stellar-to-halo specific angular momentum relation,
and one for the disc-to-virial rotation velocity relation. The shape
of the observed scaling laws carries enough information to constrain
these three quantities and their scatter together simultaneously, and to disentangle
whether a simple, linear galaxy--halo correspondence is preferred for spirals
or if a more complex, non-linear correspondence is needed \citep[e.g.][]{Lapi+18}.

In this paper we use individual, high-quality measurements of the
photometry and gas rotation velocity of a wide sample of nearby spiral
galaxies, from the smallest dwarfs to the most massive giant spirals,
to fit their observed scaling relations with analytic galaxy formation
models that depend on the three fundamental parameters mentioned above.
We perform fits of models with either i) a simple, linear galaxy--halo
correspondence, ii) a more complex, non-linear correspondence, and iii) also a
complex, non-linear correspondence that has an additional prior on the
stellar-to-halo mass relation from popular abundance matching models.
We then statistically evaluate the goodness of fit in all three cases
and, finally, we compare the outcomes of these three cases with the
halo masses recently measured from the rotation curves of the same
spirals; thus, we  have additional and independent information on which
of the models we tried is more realistic.

The paper is organised as follows. In Section \ref{sec:data} we describe
the dataset that we use; in Section \ref{sec:model} we introduce the
analytic models that we adopt to fit the observed scaling relations and
our fitting technique; in Section \ref{sec:results} we present the fitting
results, the predictions of the models, and the a posteriori comparison with the
halo masses measured from the rotation curve decompositions; in Section
\ref{sec:concl} we summarise and discuss the implications of our findings.

Throughout the paper we use a fixed critical overdensity parameter
$\Delta=200$ to define dark matter haloes and the standard $\Lambda$CDM
model, which has the following parameters estimated by the \cite{Planck18}:
$\fb\equiv\Omegab/\Omegam \simeq 0.157$ and
$H_0=67.4$ km s$^{-1}$ Mpc$^{-1}$.

\section{Data} \label{sec:data}

\subsection{SPARC}

Our primary data catalogue comes from the sample of 175 nearby disc galaxies with near-infrared
photometry and HI rotation curves (SPARC) collected by \citet[][hereafter \citetalias{SPARC}]{SPARC}.
These galaxies span more than 4 orders of magnitude in luminosity at 3.6 $\mu$m and all morphological
types, from irregulars to lenticulars. The sample was primarily collected for studies of high-quality,
regular, and extended rotation curves; thus galaxies have been primarily selected on the
basis of interferometric radio data. Moreover, the catalogue selection has been refined to include
only galaxies with near-infrared photometry from the Spitzer Space Telescope.
Hence, even though it is not volume limited, this sample provides a fair representation of the full
population of nearby (regularly rotating) spirals. Samples of spirals with a much higher
completeness and with high-quality HI kinematics will soon be available with the Square Kilometre
Array precursors and pathfinders, such as MeerKAT or APERture Tile In Focus (APERTIF).

In what follows we consider only galaxies with inclinations larger than $30^\circ$, since for nearly
face-on spirals the rotation curves are highly uncertain. This introduces no biases, since discs
are randomly orientated with respect to the line of sight.

We used the gas rotation velocity along the flat part of the rotation curve as a representative
velocity for the system because it is known to minimise the scatter of the (baryonic) Tully-Fisher
relation \citep[e.g.][]{Verheijen01,Lelli+19}.
We used the same estimate of $\vf$ as in \cite{Lelli+16a}, which is basically
an average of the three last measured points of the rotation curve, with the condition that the
curve is flat within $\sim 5\%$ over these last three points. When fitting the models in the following
sections, we only consider the sample of galaxies that satisfies this condition; this includes 125 galaxies.
We  nonetheless show the locations on the scaling relations of the other 33 galaxies
(with inclinations larger than $30^\circ$) that do not satisfy that criterion (white filled circles);
also for these objects we adopted the definition of $\vf$ and its uncertainty from \cite{Lelli+16a}.

The disc scale lengths $\rd$ have also been derived by \cite{SPARC} with exponential fits to the
outer parts of the measured surface brightness at 3.6 $\mu$m with Spitzer. These authors did this to
exclude the contamination from the bulge (if present) in the central regions of the galaxy.
We computed the stellar masses $\mstar$ by integrating the observed surface brightness profiles, which are
decomposed into a disc and bulge component as in \cite{SPARC}, and by assuming a constant
mass-to-light ratio for the two components of $(M/L_{\rm disc}^{[3/6]}, M/L_{\rm bulge}^{[3/6]})
=(0.5,0.7)$. We justified this choice by stellar population synthesis models \citep[e.g.][]
{SchombertMcGaugh14} and is found to minimise the scatter of the (baryonic) Tully-Fisher
\citep{Lelli+16a,Ponomareva+18}. Moreover, these values are similar to those obtained
from the mass decomposition of the rotation curves (\citealt{Katz+17}, \citetalias{PFM19}).

The $\jstar-\mstar$ relation, aka the Fall relation, is now very well established
observationally. Several independent measurements now agree perfectly both
on the slope and normalisation of this relation at least for spirals
\citep{RF12,ObreschkowGlazebrook14,Posti+18b,FR13,FR18}.
The total specific angular momentum of the stellar disc is, instead, measured as in
\cite{Posti+18b}. Given the stellar rotation curve $V_\star$, estimated from the HI rotation
curve\footnote{
After accounting for the support from the stellar velocity dispersion, or the so-called
asymmetric drift correction, following the measurements from \cite{Martinsson+13}.
This correction is found to be negligible for the determination of $\jstar$ for most
systems \citep{Posti+18b}.
}, and the stellar surface density $\Sigma_\star$, we calculated
\begin{equation} \label{def:jstar}
    \jstar = \frac{\int\de R\,R^2\,\Sigma_\star(R)\,V_\star(R)}{\int\de R\,R\,\Sigma_\star(R)}.
\end{equation}
We used this measurement \citep[and associated uncertainty as given by Eq. 3 in][]{Posti+18b}
for the 92 SPARC galaxies with ``converged'' cumulative $\jstar$ profiles, meaning that
they flatten in the outskirts to within $\sim 10\%$ \citep[following the definition by][]
{Posti+18b}. For the other 33 galaxies with flat rotation curves, but with non-converged
cumulative $\jstar$ profiles, we adopted the much simpler estimator \citep[see e.g.][]{RF12}
\begin{equation} \label{def:jstarRF12}
    \jstar = 2\,\rd\,\vf,
\end{equation}
which comes from Eq.~\eqref{def:jstar} under the assumption of an exponential stellar
surface density profile with a flat rotation curve. In this equation, we are implicitly assuming that
the gas rotation, $\vf$, is a reasonable proxy for the rotation velocity of stars, at
least in the outer regions of discs. Stars are indeed found on almost circular
orbits in the regularly rotating discs analysed in this work \citep{Iorio+17,Posti+18b}.
The simple $\jstar$ estimator in Eq.~\eqref{def:jstarRF12} is widely used and
known to be reasonably accurate for spirals, provided that the measurements of $\rd$ and
$\vf$ are sound \citep[e.g.][]{FR18}. In particular, \cite{Posti+18b},
studying the sample of 92 SPARC galaxies with converged profiles, determined that
the estimator \eqref{def:jstarRF12} is unbiased and yields a typical uncertainty
of $30-40\%$ on the true $\jstar$. Thus, in what follows, we also consider $\jstar$
measurements obtained with Eq.~\eqref{def:jstarRF12} and with an uncertainty of $40\%$
for the 33 SPARC galaxies with flat rotation curves, but non-converged cumulative
$\jstar$ profiles.

\subsection{LITTLE THINGS}
We added a sub-sample of galaxies drawn from the Local Irregulars That Trace Luminosity
Extremes, The HI Nearby Galaxy Survey Survey \citep[LITTLE THINGS,][]{Hunter+12} to
the catalogue described above. These are 17 dwarf irregulars that have fairly regular HI
kinematics and are seen at inclinations larger than $30^\circ$.

This sample has been recently analysed by \citet{Iorio+17} who determined the rotation
curve of each system from the detailed 3D modelling of the HI data.
We used their results and applied the same criterion on the rotation curve flatness as for
the SPARC sample. We found that 4 out of 17 galaxies (CVnIdwA, DDO53, DDO210, UGC8508)
have rotation curves which do not flatten to within $\sim 5\%$ over the last three data
points, and thus we excluded these galaxies from the fits but we still show these in the plots (as white
filled diamonds).

We determined the size of these galaxies from their optical R-band or V-band images using
publicly available data from 1-2 meter class telescopes at the Kitt Peak National
Observatory \citep[KPNO;][]{Cook+14}.
In the cases where no KPNO data were available, we used Sloan Digital Sky Survey (SDSS)
data \citep[CVnIdwA, DDO\,101, DDO\,47, and DDO\,52;][]{Baillard+11} or Vatican Advanced
Technology Telescope (VATT) data \citep[UGC\,8508;][]{Taylor+05} instead.
While a number of LITTLE THINGS systems come with IRAC Spitzer images, these are vastly
contaminated by bright point-like sources that we found difficult to treat properly.
Also considering the superior quality of the optical data, we decided to use the latter
for our size measurements.

\begin{figure}
\includegraphics[width=0.5\textwidth]{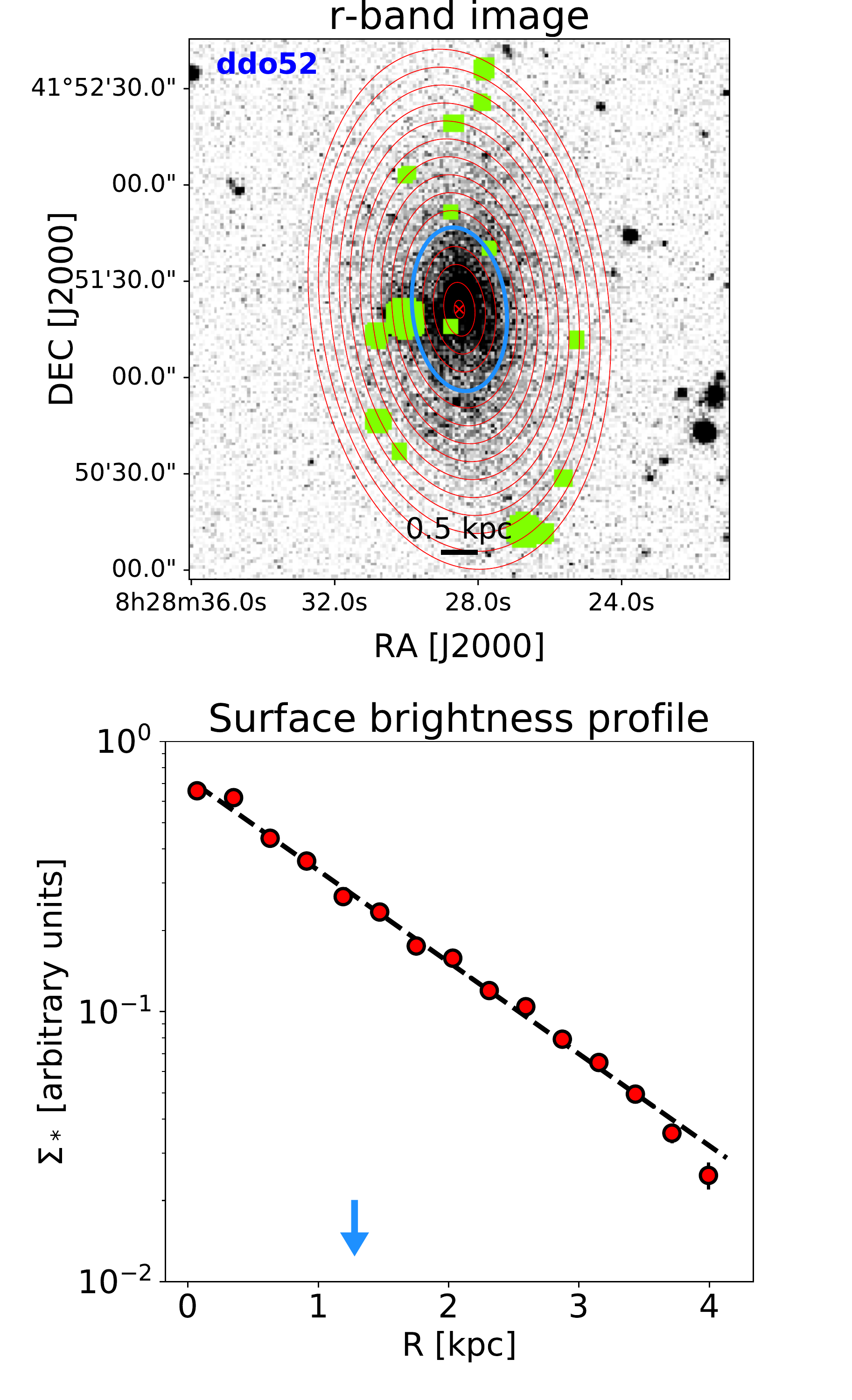}
\caption{Photometry for DDO\,52 as a representative example for the LITTLE THINGS
         galaxies. Top panel: r-band image with the concentric ellipses showing
         the annuli where the surface brightness is computed. The green regions are
         foreground sources that we mask during the derivation of the profile.
         The blue ellipse is drawn at the disc scale length.
         Bottom panel: surface brightness profile, normalised to the total light
         within the outermost ring. The thick dashed black line indicates the exponential fit,
         while the blue arrow indicates the exponential scale length.
        }
\label{fig:LT_phot}
\end{figure}

Using these images, we derived the surface brightness profiles for all 17 systems following
the procedure fully described in \citet{Marasco+19}, adopting as galaxy centres, inclinations
and position angles the values determined by \citet{Iorio+17}.
We then fit these profiles with exponential functions to determine the galaxy scale lengths,
which we found to be in excellent agreement with those inferred by \citet{HunterElmegreen06}.
In Figure~\ref{fig:LT_phot} we illustrate the procedure we use for the representative case of
DDO\,52.
Finally, for the LITTLE THINGS galaxies we used the estimator \eqref{def:jstarRF12} for the
stellar specific angular momentum with a conservative error bar of $40\%$.

\section{Model} \label{sec:model}

\subsection{Dark matter haloes}

We started with dark matter haloes, which are described by their structural
properties -- mass ($\Mh$), radius ($\Rh$), velocity ($\Vh$), and specific
angular momentum ($\jh$) -- defined in an overdensity of $\Delta$ times the critical
density of the Universe. Haloes, then, adhere to the following scaling laws
\citep[e.g.][]{MvdBW10}:
\begin{equation} \label{eq:dm_mv}
    \Mh = \frac{1}{GH}\sqrt{\frac{2}{\Delta}} \Vh^3;
\end{equation}
\begin{equation} \label{eq:dm_rv}
    \Rh = \frac{1}{H}\sqrt{\frac{2}{\Delta}} \Vh;
\end{equation}
\begin{equation} \label{eq:dm_jv}
    \jh = \frac{2\lambda}{H\sqrt{\Delta}} \Vh^2,\end{equation}
where $G$ is the gravitational constant and $\lambda = \jh / \sqrt{2} \Rh\Vh$ is
the halo spin parameter, as in the definition by \citet[][which is conceptually
equivalent to the classic definition in \citealt{Peebles69}]{Bullock+01}. The distribution
of $\lambda$ for $\Lambda$CDM haloes is very well studied and it is known to have
a nearly log-normal shape -- with mean $\log\overline{\lambda} \approx -1.456$ and
scatter $\sigma_{\log\lambda}\approx 0.22$ dex -- irrespective of halo mass.
Henceforth, since $\lambda$ is not a function of $\Vh$, Eq.~\eqref{eq:dm_jv} is
a simple power law $\jh\propto\Vh^2$, while also Eq.~\eqref{eq:dm_mv}-\eqref{eq:dm_rv}
are obviously similar power laws.

\subsection{Galaxy formation parameters}

We very simply parametrise the intrinsically complex processes of galaxy formation,
by considering that, to first order, galaxies form out of the cooling of baryons
within the virial radius of their halo. The fundamental parameters we consider are
then the following fractions:
\begin{equation} \label{def:fracs}
    \fM\equiv\frac{\mstar}{\Mh}; \qquad \fj\equiv\frac{\jstar}{\jh}; \qquad
    \fv\equiv\frac{\vf}{\Vh}; \qquad \fr\equiv\frac{\rd}{\Rh}.
\end{equation}
The aim of this work is to unveil the galaxy--halo connection constraining and
characterising the four galaxy formation fractions above  using the observed
global properties of disc galaxies.

\begin{itemize}

\item The first, and arguably most important, of these parameters is the stellar mass
fraction $\fM$, which is also sometimes loosely referred to as global star formation
efficiency parameter (e.g. \citealt{Behroozi+13}; \citetalias{Moster+13}).
This describes how much of the hot gas associated with the dark matter halo was
able to cool and to form stars throughout the lifetime of the galaxy. Thus, in a very
broad sense, this encapsulates the average efficiency of gas-to-stars conversion
integrated over time. On average, this parameter has an obvious strict upper limit
set by the cosmic baryon fraction $\fb\simeq 0.157$.

\item The second is the specific angular momentum fraction $\fj$, also known as the
retained fraction of angular momentum \citep[e.g.][]{RF12}.
After the halo collapsed, but before galaxy formation started, tidal torques
supplied baryons and dark matter with nearly equal amounts of angular momentum,
so $\jbar/\jh=1$ \citep[e.g.][]{FallEfstathiou80}.
Stars, however, form from some fraction of these baryons, whose final angular
momentum distribution is the product of the interplay of several physical
processes (e.g. cooling, mergers, and feedback, gas cycles). This results in an
$\fj$ that can easily deviate from unity.

\item The third is the velocity fraction $\fv$, which is the ratio of the circular
velocity at the edge of the galactic disc to that at the virial radius. While
this factor in principle can take any value depending on the galaxy and halo mass
distribution, but also depending on the extension of the measured rotation curve,
it is typically expected to be $\fv \gtrsim 1$ for large and well resolved galaxies
\citep[$\log\mstar/\Msun > 9$, see e.g.][]{Papastergis+11}.

\item The last parameter is the size fraction $\fr$, i.e. the ratio of the disc
exponential scale length ($\rd$) to the halo virial radius ($\Rh$). However,
if we assume that the size of the galaxy is regulated by its angular momentum
\citep{FallEfstathiou80,MMW98, Kravtsov13}, then $\fj$ and $\fr$ are not
independent. It is easy to work out their relation as a function of the
dark matter halo profile, which turns out to be analytic in the case of an
exponential disc with a flat rotation curve (see Appendix~\ref{app:fr}), i.e.,
\begin{equation}
    \fr=\frac{\lambda}{\sqrt{2}}\frac{\fj}{\fv}.
\end{equation}
An analogous result was already derived analytically by \cite{Fall83}.
For more realistic haloes, for example a \citet[][NFW]{NFW} halo, a similar
proportionality still exists, and can be worked out with an iterative procedure
\citep[see e.g.][]{MMW98}.
\end{itemize}

With these definitions we can rewrite the dark matter relations of
Eqs.~\eqref{eq:dm_mv}-\eqref{eq:dm_jv} now for  the stellar discs as
\begin{equation} \label{eq:stars_mv}
    \mstar = \frac{\fM}{GH}\sqrt{\frac{2}{\Delta}} \left(\frac{\vf}{\fv}\right)^3;
\end{equation}
\begin{equation} \label{eq:stars_rv}
    \rd = \frac{\lambda\fj}{H\sqrt{\Delta}} \frac{\vf}{\fv^2};
\end{equation}
\begin{equation} \label{eq:stars_jv}
    \jstar = \frac{2\lambda\fj}{H\sqrt{\Delta}} \left(\frac{\vf}{\fv}\right)^2.
\end{equation}
In this form, the above equations involve all observable quantities ($\vf,\mstar,\rd,
\jstar$) and the three fundamental fractions ($\fM,\fj,\fv$).
In what follows, we use observations on the $\rd-\vf$ and the $\jstar-\vf$ diagrams,
together with the usual stellar mass Tully-Fisher $\mstar-\vf$, instead of the
more canonical size-mass and Fall relations. The main reason for this is that when
high-quality HI interferometric data are available, $\vf$ is a very well-measured
quantity (typically within $\sim 5\%$), while $\mstar$ suffers from many systematic
uncertainties (e.g. on the stellar initial mass function). Thus, we use the observed
scaling relations \eqref{eq:stars_mv}-\eqref{eq:stars_jv} to constrain the behaviour
of the three fundamental fractions as a function of $\vf$. However, we show in
Appendix~\ref{app:mstar} the result of fitting the more canonical Tully-Fisher,
size-mass, and Fall relations, hence deriving the fractions \eqref{def:fracs} as a
function of $\mstar$. We note that, as might be expected, we find similar results
for the fractions $\fM$, $\fj$, and $\fv$ when having either $\vf$ or $\mstar$ as the
independent variable for the scaling laws.

\subsection{Functional forms of the fractions $\fM$, $\fj$, and $\fv$} \label{sec:fractions}

The three scaling laws \eqref{eq:stars_mv}-\eqref{eq:stars_jv} provide us with constraints
on the three fundamental galaxy formation parameters $\fM$, $\fj$, and $\fv$. In
particular, these are generally not constant (e.g. \citetalias{Moster+13} for $\fM$;
\citealt{Posti+18a} for $\fj$; \citealt{Papastergis+11} for $\fv$) and their variation
from dwarf to massive galaxies is  encoded in the scaling laws.
We use parametric functions to describe the behaviour of $\fM$, $\fj$, and $\fv$ as a
function of $\vf$ and then we look for the parameters that yield the best match to the
observed scaling relations. The ansatz on the functional form of
$f=f(\vf)$, where $f$ is any of the three fractions, and the prior knowledge imposed
on some of the free parameters, define the three models that we test in this paper.

\begin{itemize}
    \item[\bf (i)] In the first and simplest model that we consider, the three fractions
    $\log\,f$ to vary linearly as a function of $\log\,\vf$ as follows:
    \begin{equation} \label{eq:lin}
        \log\, f = \alpha\log\,\vf/{\rm km\,s^{-1}}+\lfzero.
    \end{equation}
    Thus, we have a slope ($\alpha$) and a normalisation ($\fzero$) for each of the three
    fractions $\fM$, $\fj$, and $\fv$. In this case, we adopt uninformative priors for all
    the free parameters.\newline

    \item[\bf (ii)] The second model assumes a more complicated double power-law dependence
    of $f$ on $\vf$,
    \begin{equation} \label{eq:dpl}
        f = \fzero \left(\frac{\vf}{\vzero}\right)^\alpha
        \left(1+\frac{\vf}{\vzero}\right)^{\beta-\alpha}
    .\end{equation}
    We have two slopes ($\alpha$, $\beta$) and a normalisation ($\fzero$) that are
    different for each of the three $f$; while the scale velocity ($\vzero$), which defines
    the transition between the two power-law regimes, is the same for the three fractions for
    computational simplicity.
    Also in this case, we use uninformative priors for all the free parameters.\newline

    \item[\bf (iii)] The last model has the same functional form as model (ii), i.e.
    Eq.~\eqref{eq:dpl}, with uninformative priors for $\fj$ and $\fv$; while we impose
    normal priors on the slopes ($\alpha$, $\beta$), normalisation ($\fzero$) and scale velocity
    ($\vzero$) such that the global star formation efficiency follows the results
    of the abundance matching model by \citetalias{Moster+13}. In order to properly account
    for the sharp maximum of $\fM$ at $\Mh\approx 4\times 10^{11}\Msun$, we slightly modify
    the functional form of $\fM=\fM(\vf)$ as
    \begin{equation} \label{eq:m13prior}
        \fM = \fzero \left(\frac{\vf}{\vzero}\right)^\alpha
        \left[1+\left(\frac{\vf}{\vzero}\right)^\gamma\right]^{\beta-\alpha},
    \end{equation}
    where $\gamma=3$ since $\Mh\propto\vf^3$.
\end{itemize}

While the ansatz (i) was chosen because it is the simplest possible, with the smallest
number of free parameters, the functional form and priors adopted in cases (ii) and (iii)
were inspired by many results obtained using different methods on the stellar-to-halo mass
relation \citep[see][and references therein]{WechslerTinker18}. Thus, in case (ii) we allow
$\fM$, but also $\fj$ and $\fv$, to follow the double power-law functional form, which is
typically used to parametrise how $\fM$ varies for galaxies of different masses; while in
case (iii) we additionally impose priors on the $\fM$ parameters, following the results
of one of the most popular stellar-to-halo mass relations \citepalias{Moster+13}.

In both models (ii) and (iii), the scale velocity $\vzero$ is the only parameter that
we assume to be the same for $\fM$, $\fj$, and $\fv$. The reason is mainly statistical,
as the data are not informative enough to disentangle between breaks occurring at
different $\vzero$ for different fractions. The observed scaling relations
carry enough statistical information to distinguish only basic trends (for instance,
whether or not there is a peak in $\fM$, $\fj$, and/or $\fv$) and cannot
really discriminate between detailed, degenerate behaviours. Moreover, both $\fj$
and $\fv$ are thought to be physically, closely related to $\fM$
\citep[e.g.][]{NavarroSteinmetz00,Cattaneo+14,Posti+18a}, so it makes sense to
investigate a scenario in which they have a transition at the same physical galaxy
mass scale. In what follows, we dub the models (i)-(ii)-(iii) as linear,
double power law and \citetalias{Moster+13} prior, respectively.

Finally, we note that we also tried letting free the parameter governing the sharpness
of the transition of the two power-law regimes; i.e. $\gamma$ in Eq.~\eqref{eq:m13prior}.
Again, we find that the data do not have enough information to constrain this variable,
thus we decided to fix it to $\gamma=1$ (as in the double power-law model). Fixing
it to other values (e.g. $\gamma=3$, as in the \citetalias{Moster+13} prior model)
yields similar results to those presented below.

\subsection{Intrinsic scatter} \label{sec:scatter}

In all models we allow the three fractions $\fM$, $\fj$, and $\fv$ to have a non-null
intrinsic scatter $\sigma$. This parameter has an important physical meaning, as it
encapsulates all the physical variations of the complex processes that lead to the
formation of galaxies. The information on this parameter comes from the intrinsic
vertical scatter (at fixed $\vf$) observed in the three different scaling relations
considered in this work. All of the measured scatters $\siglm$, $\siglr$, and $\siglj$ are
given by the combination of the intrinsic scatter of $\fv$ with that of $\fM$ or
$\fj$. This combination is clearly degenerate and the information encoded in the
data is not enough to distinguish the two of them\footnote{
Indeed, we tried letting free both the scatter of $\fv$ and that of $\fM$ or $\fj$
finding a non-flat posterior in only one of the two, which happens to be compatible
with the value we quote in Tab.~\ref{tab:parameters}}.
Henceforth, for simplicity we assume that the intrinsic scatter is the same for
all three fractions.

With this simplifying assumption, the scatter on $\log\,f$ ($\siglf$) is related to
the observed intrinsic vertical scatter of the three scaling relations as
\begin{equation} \label{eq:scat_m}
    \siglm = \sqrt{10}\,\siglf,
\end{equation}
\begin{equation} \label{eq:scat_r_j}
    \siglr = \siglj = \sqrt{5\siglf^2+\siglam^2},
\end{equation}
where $\siglam\approx 0.22$ dex is the known scatter on the halo spin parameter.
These formulae come from standard propagation of uncertainties in Eqs.
\eqref{eq:stars_mv}-\eqref{eq:stars_jv}, where only the non-null intrinsic scatters
of the fractions $f$ and the halo spin parameter $\lambda$ are considered.
 An additional free parameter in every model we tried is $\siglf$; thus, all in all,
model (i) has 7 free parameters, while models (ii)-(iii) have 11 free parameters.

\subsection{Likelihood and model comparison} \label{sec:like}

We use Bayesian inference to derive posterior probabilities of the free parameters
($\boldsymbol{\theta}$) in our three sets of models, i.e.
\begin{equation} \label{eq:bayes}
    \mathcal{P}(\boldsymbol{\theta}|\vf,\mstar,\rd,\jstar)\propto
        \mathcal{P}(\vf,\mstar,\rd,\jstar|\boldsymbol{\theta})\,
        \mathcal{P}(\boldsymbol{\theta}),
\end{equation}
where $(\vf,\mstar,\rd,\jstar)$ are the data, $\mathcal{P}(\boldsymbol{\theta})$ is
the prior, and $\mathcal{P}(\vf,\mstar,\rd,\jstar|\boldsymbol{\theta})$ is the likelihood.
The prior is uninformative (flat) for all free parameters, except in model (iii)
where it is normal for the four parameters describing $\fM$ where means and standard
deviations have been taken from the abundance matching model of \citetalias{Moster+13}.
The likelihood is defined as a sum of standard $\chi^2$, i.e.
\begin{equation} \label{eq:loglike}
        \ln{\mathcal{P}(\vf,\mstar,\rd,\jstar|\boldsymbol{\theta})} =
        \ln{\mathcal{P}_M}+\ln{\mathcal{P}_R}+\ln{\mathcal{P}_j},
\end{equation}
where
\begin{equation} \label{eq:loglikeMstar}
        \ln{\mathcal{P}_M} = -\sum_{i=0}^{N} \frac{1}{2}\frac{\left[\mstar -
        \mstar(\vf)^{\rm Eq.\eqref{eq:stars_mv}}\right]^2}{\siglm^2+\delta_{\mstar}^2} -
        \frac{1}{2}\log\left[2\pi\left(\siglm^2+\delta_{\mstar}^2\right)\right],
\end{equation}
\begin{equation} \label{eq:loglikeRd}
        \ln{\mathcal{P}_R} = -\sum_{i=0}^{N} \frac{1}{2}\frac{\left[\rd -
        \rd(\vf)^{\rm Eq.\eqref{eq:stars_rv}}\right]^2}{\siglr^2+\delta_{\rd}^2} -
        \frac{1}{2}\log\left[2\pi\left(\siglr^2+\delta_{\rd}^2\right)\right],
\end{equation}
\begin{equation} \label{eq:loglikej}
        \ln{\mathcal{P}_j} = -\sum_{i=0}^{N} \frac{1}{2}\frac{\left[\jstar -
        \jstar(\vf)^{\rm Eq.\eqref{eq:stars_jv}}\right]^2}{\siglj^2+\delta_{\jstar}^2} -
        \frac{1}{2}\log\left[2\pi\left(\siglj^2+\delta_{\jstar}^2\right)\right],
\end{equation}
and $\delta_{\mstar}$, $\delta_{\rd}$, and $\delta_{\jstar}$ are the measurement
uncertainties on the respective quantities. We note that this likelihood does
not account for the observational uncertainties on $\vf$, which are much smaller
than those on the other observable quantities. This implies that the intrinsic scatter
$\siglf$ that we fit is vertical and that it is greater or equal to the
intrinsic perpendicular scatter.

Given these definitions, we construct the posterior
$\mathcal{P}(\boldsymbol{\theta}|\vf,\mstar,\rd,\jstar)$ with a Monte Carlo Markov Chain
method \citep[MCMC; and in particular with the \texttt{python} implementation by][]
{emcee}. In each of the three cases (i)-(ii)-(iii), we define the ``best model'' to
be the model that maximises the log-likelihood.

Finally, we assess which one between the three best models is preferred by the data
using standard statistical information criteria: the Akaike information criterion (AIC)
and Bayesian information criterion (BIC). These are meant to find the best
statistical compromise between goodness of fit (high $\ln\mathcal{P}$) and model complexity
(less free parameters), in such a way that any gain in having a larger likelihood is
penalised by the amount of new free parameters introduced. The preferred model is then
chosen as that with the smallest AIC and BIC amongst those explored.

\begin{table}
\caption{Posterior distributions of the parameters of the three models considered in
this study. The three columns are for the linear (Eq.~\ref{eq:lin}),
double power law (Eq.~\ref{eq:dpl}) and \citetalias{Moster+13} prior models,
respectively (Eq.~\ref{eq:m13prior}). The four row blocks, instead, refer to
the retained fraction of angular momentum $\fj$, the star formation efficiency
$\fM$, the ratio of asymptotic-to-virial velocity $\fv$, and their intrinsic
scatter $\siglf$. The posteriors of the parameters are all summarised with
their 16th-50th-84th percentiles.}
\label{tab:parameters}
\begin{center}
\setlength\extrarowheight{5pt}
\begin{tabular}{@{\extracolsep{\fill}} lr|r|r}
& linear & double power-law & \citetalias{Moster+13} prior\\
\hline\hline
 $\log\,f_{0,j}$ & $-0.33^{+0.39}_{-0.40}$ & $0.03^{+0.48}_{-0.50}$ & $2.59^{+0.37}_{-0.37}$ \\[5pt]
 $V_0/{\rm km\,s^{-1}}$ & -- & $63000^{+300000}_{-45000}$ & $124^{+8}_{-8}$ \\[5pt]
 $\alpha_{j}$ & $0.08^{+0.19}_{-0.19}$ & $0.1^{+0.16}_{-0.17}$ & $4.1^{+0.5}_{-0.5}$ \\[5pt]
 $\beta_{j}$ & -- & $1^{+31}_{-30}$ & $-4.4^{+0.7}_{-0.7}$ \\[5pt]
\hline
 $\log\,f_{0,M}$ & $-5.07^{+0.43}_{-0.43}$ & $2.01^{+1.07}_{-1.02}$ & $-1.01^{+0.07}_{-0.07}$ \\[5pt]
 $\alpha_{M}$ & $1.46^{+0.21}_{-0.21}$ & $-1.45^{+0.26}_{-0.21}$ & $4.33^{+0.09}_{-0.09}$ \\[5pt]
 $\beta_{M}$ & -- &$0^{+33}_{-30}$ & $2.22^{+0.09}_{-0.09}$ \\[5pt]
\hline
 $\log\,f_{0,V}$ & $0.04^{+0.13}_{-0.13}$  & $0.16^{+0.20}_{-0.19}$  & $1.13^{+0.12}_{-0.12}$ \\[5pt]
 $\alpha_{V}$ & $0.01^{+0.06}_{-0.06}$ & $0.05^{+0.08}_{-0.06}$ & $1.6^{+0.2}_{-0.2}$ \\[5pt]
 $\beta_{V}$ & -- & $-15^{+18}_{-23}$ & $-1.8^{+0.2}_{-0.2}$ \\[5pt]
\hline
 $\sigma_{\log\,f}$ & $0.07^{+0.01}_{-0.01}$ & $0.07^{+0.01}_{-0.01}$ & $0.08^{+0.01}_{-0.01}$\\[5pt]
\end{tabular}
\end{center}
\end{table}

\begin{table}
\caption{Goodness of fit of the three best models.}
\label{tab:fits}
\begin{center}
\begin{tabular}{lcccc}
\hline\hline \\[-.2cm]
Model & $\ln\mathcal{P}_{\rm max}$ & $\Delta$ AIC &  $\Delta$ BIC  \vspace{.1cm}\\
\hline\hline \\[-.2cm]
linear                       & $-39.4$ & $0$    & $0$    \\
double power-law             & $-37.5$ & $3.2$  & $18.4$ \\
\citetalias{Moster+13} prior & $-61.9$ & $23.4$ & $52.1$ \\
\hline
\end{tabular}
\end{center}
\end{table}

\section{Results} \label{sec:results}
\subsection{Fits of the scaling laws and model comparison}

\begin{figure*}
\includegraphics[width=\textwidth]{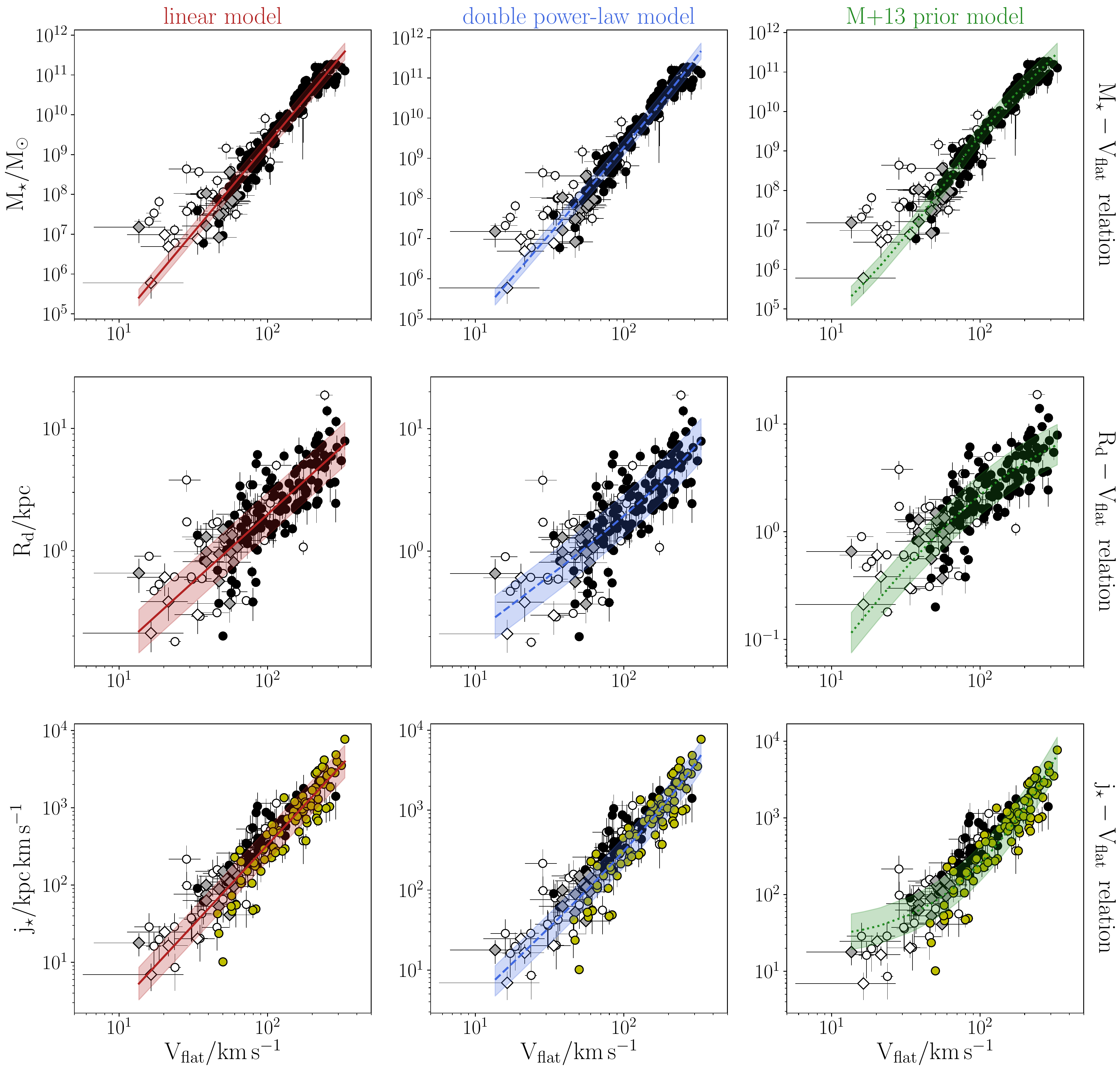}
\caption{Comparison of the three best models obtained with different assumptions on
         $\fj$, $\fM$, and $\fv$ with the data from SPARC galaxies (circles) and
         LITTLE\ THINGS galaxies (diamonds). Each column shows the fits
         for a given model, following the assumptions in Sec.~\ref{sec:fractions}.
         The top row is for the stellar Tully-Fisher relation, the middle row is for the
         size-velocity relation, while the bottom row is for the angular
         momentum-velocity relation. The white filled points in the plots are the galaxies
         which do not satisfy the \cite{Lelli+16a} criterion on the flatness of their
         rotation curves. The yellow filled points in the $\jstar-\vf$
         relation are the 92 SPARC galaxies with `onverged $\jstar$ profiles,
         following \cite{Posti+18b}.
        }
\label{fig:fits}
\end{figure*}

We modelled the observed $\mstar-\vf$, $\rd-\vf$ and $\jstar-\vf$ relations with
the three models described in Sec.~\ref{sec:fractions}. We have found the
best model, defined as the maximum a posteriori, in the three cases (i)-(ii)-(iii)
and we show how they compare with the observations in Figure \ref{fig:fits}.
In each row of this Figure we show one of the three scaling relations
considered; while in each column we present the comparison of the data
with the three models.
Table~\ref{tab:parameters} summarises the posterior distributions that we derive for
the parameters of the three models (with their 16th-50th-84th percentiles).

We only fitted the data for galaxies which have a flat rotation curve according to the
definition in Sec.~\ref{sec:data} (i.e. black-, grey- and gold-filled points).
The first noteworthy result is that all three best models provide a reasonably good
description of the observed nearby disc galaxy population. The agreement between
the distribution of the data and the predictions of the models is remarkable in all panels,
except perhaps in the $\jstar-\vf$ plane where the best \citetalias{Moster+13} prior
model seems to favour slightly higher angular momentum dwarfs than observed.
We report in Table~\ref{tab:fits} the values of the maximum-likelihood models in
the three cases.

While the general trend of stellar mass, size, and specific angular momentum as a
function of $\vf$ is well captured by the three best models, the inferred intrinsic
vertical scatters of the three scaling relations are also well reproduced.
While we measure a vertical scatter of $0.21$, $0.22,$ and $0.23$
dex for the observed $\mstar-\vf$, $\rd-\vf$ and $\jstar-\vf$ relations, respectively (with a typical
uncertainty of 0.03 dex); the vertical intrinsic scatters of the three scaling laws
predicted by the three best models (with Eqs.~\ref{eq:scat_m}-\ref{eq:scat_r_j}) are written as
\begin{equation*}
    (\siglm, \,\siglr, \,\siglj) = (0.22, 0.26, 0.26); \quad \mbox{\footnotesize(linear)}
\end{equation*}
\begin{equation*}
    (\siglm, \,\siglr, \,\siglj) = (0.21, 0.26, 0.26); \quad \mbox{\footnotesize(double power law)}
\end{equation*}
\begin{equation*}
    (\siglm, \,\siglr, \,\siglj) = (0.23, 0.27, 0.27); \quad \mbox{\footnotesize(M+13 prior)}
\end{equation*}
with a typical uncertainty of 0.02 dex.
The scatter of $\mstar$ in the models perfectly matches the observed scatter, while it
is slightly larger for $\rd$ and $\jstar$ albeit being consistent within the uncertainties.
It is well known  that the observed scatter on $\jstar$ for galactic discs is smaller
than that expected only from the distribution of halo spin parameters \citep{RF12}, which
already suggests that the intrinsic scatter of $\fj$ has to be particularly small and also
that the scatter of $\lambda$ likely correlates with that of other properties of the
galaxy--halo connection \citep[e.g. $\fj$ or $\fv$; see][]{Posti+18a}.
Interestingly despite the differences in the three models of Sec.~\ref{sec:fractions},
we find consistently in all cases that the preferred value of the intrinsic scatter on
the three fundamental fractions $\fM$, $\fj$, and $\fv$ is $\siglf=0.07 \pm 0.01$ dex.
This small scatter indicates that the galaxy--halo connection is extremely tight
in disc galaxies, independently of their complex formation process. The connection with
baryons is likely to be even tighter than with stars, as hinted
by the very small scatter of the baryonic Tully-Fisher relation. This means that
studying the observed baryonic fractions instead of stellar fractions should be
particularly illuminating in the future.

Of the three best models that we have found, the double power law model has the highest
likelihood.  This is not surprising, as this model has the
most freedom to adapt to the observed data. Employing the statistical criteria of both
AIC and BIC, it turns out that the gain in a larger value of the likelihood does not
statistically justify the inclusion of four more free parameters with respect to the
linear model. On the other hand, the \citetalias{Moster+13} prior model is by
far the least preferred by our analysis, since it has the lowest likelihood and a large
AIC and BIC difference with respect to the linear model. Thus, we have to conclude
that to fit the current observations of the scaling laws of nearby discs, any model
more complex than a single power law statistically results in an overfit.
These results are summarised in Tab.~\ref{tab:fits}.

\begin{figure*}
\includegraphics[width=\textwidth]{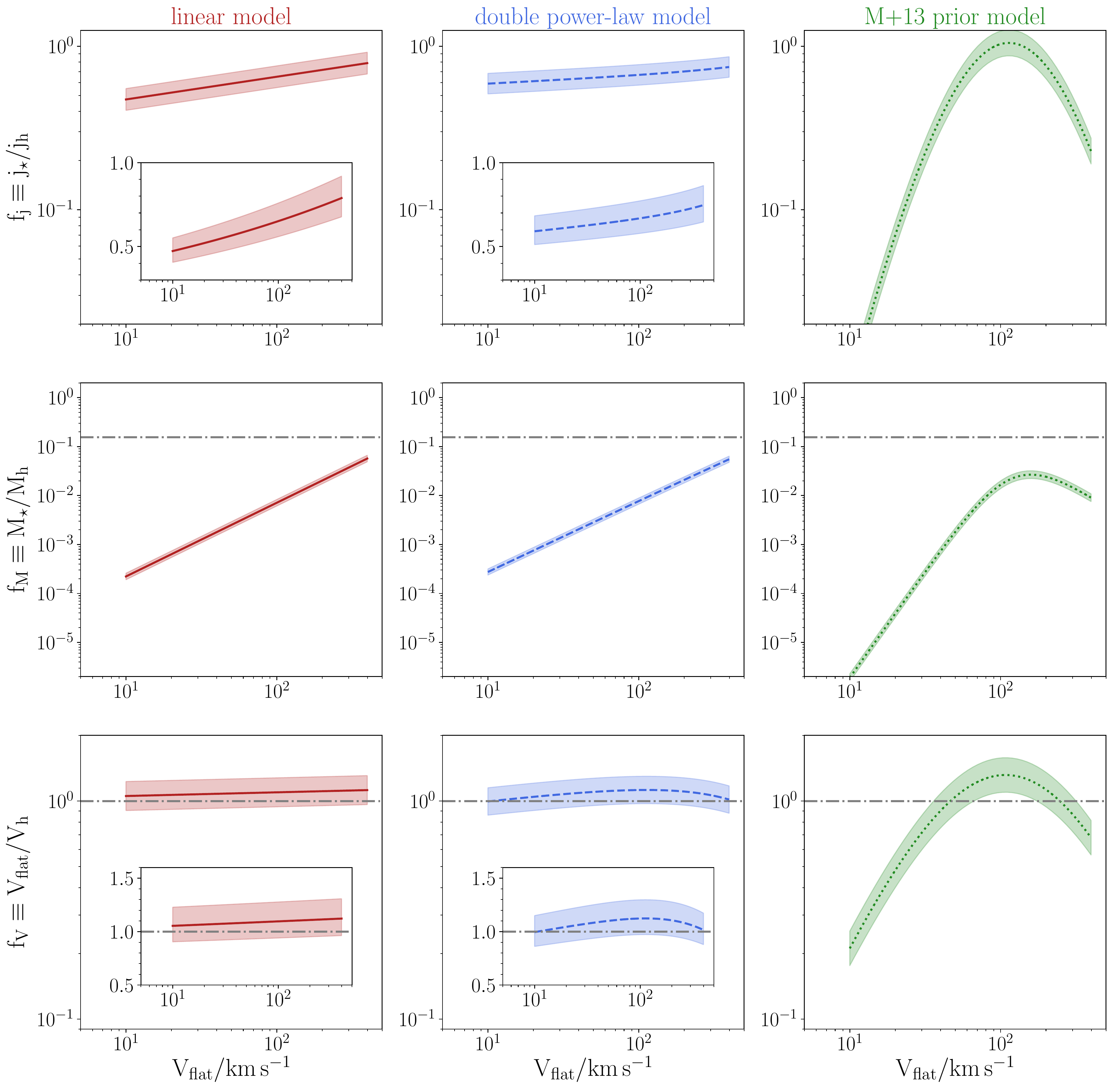}
\caption{Model predictions. We show how $\fj$ (top row), $\fM$ (middle row), and $\fv$
         (bottom rows) vary as a function of $\vf$ for the three best models (columns).
         In the middle row the dot-dashed line shows the value of the cosmic baryon
         fraction $\fb=0.157$, while in the bottom row the dot-dashed line indicates the value $\fv=1$.
         The insets show a zoom-in of the plots in linear scale.
        }
\label{fig:predictions}
\end{figure*}

\begin{figure}
\includegraphics[width=0.45\textwidth]{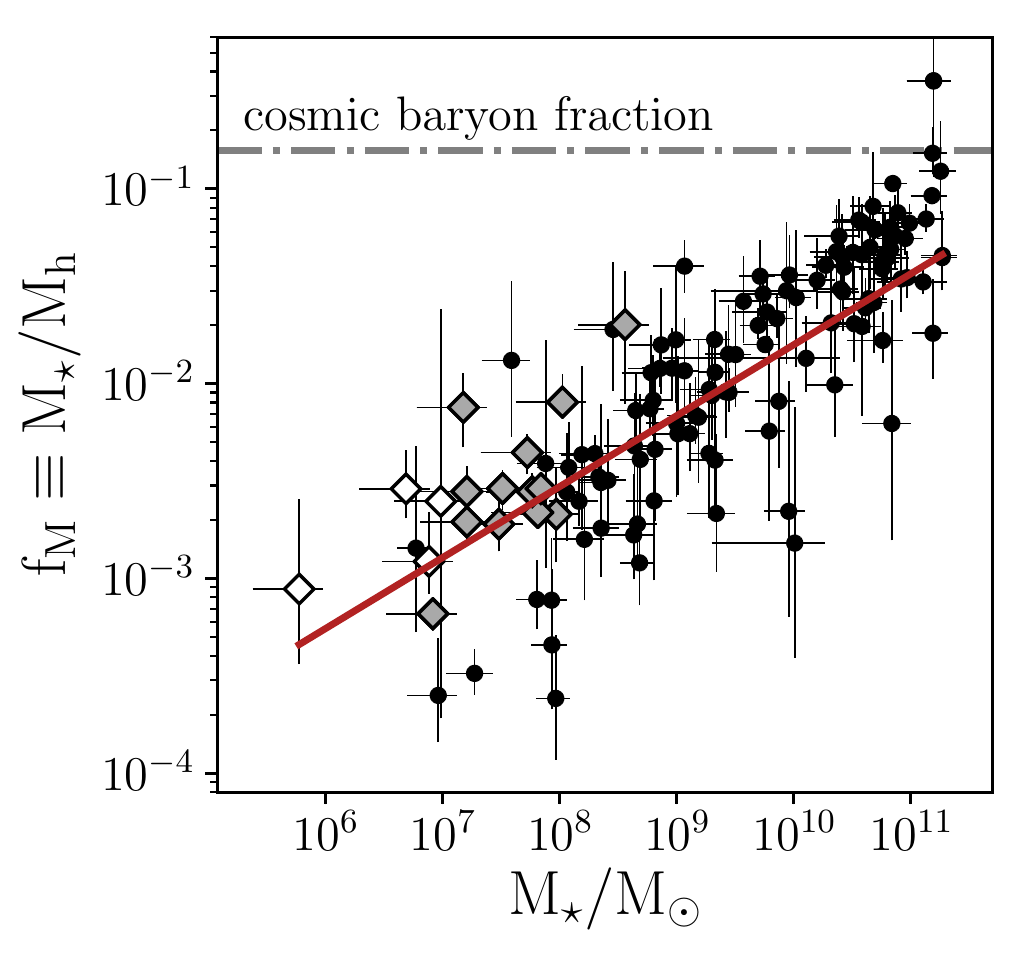}
\caption{Global star formation efficiency $\fM\equiv\Mstar/\Mh$ as a function of $\Mstar$
         for the SPARC and LITTLE THINGS galaxies. The measurements of the halo masses
         come from \citetalias{PFM19} and \citet{Read+17}, respectively.
         Symbols are the same as in Fig.~\ref{fig:fits}.
         The red line indicates the $\fM-\Mstar$ relation derived in the linear model
         for guidance.
        }
\label{fig:fM-Mstar}
\end{figure}

\begin{figure*}
\includegraphics[width=\textwidth]{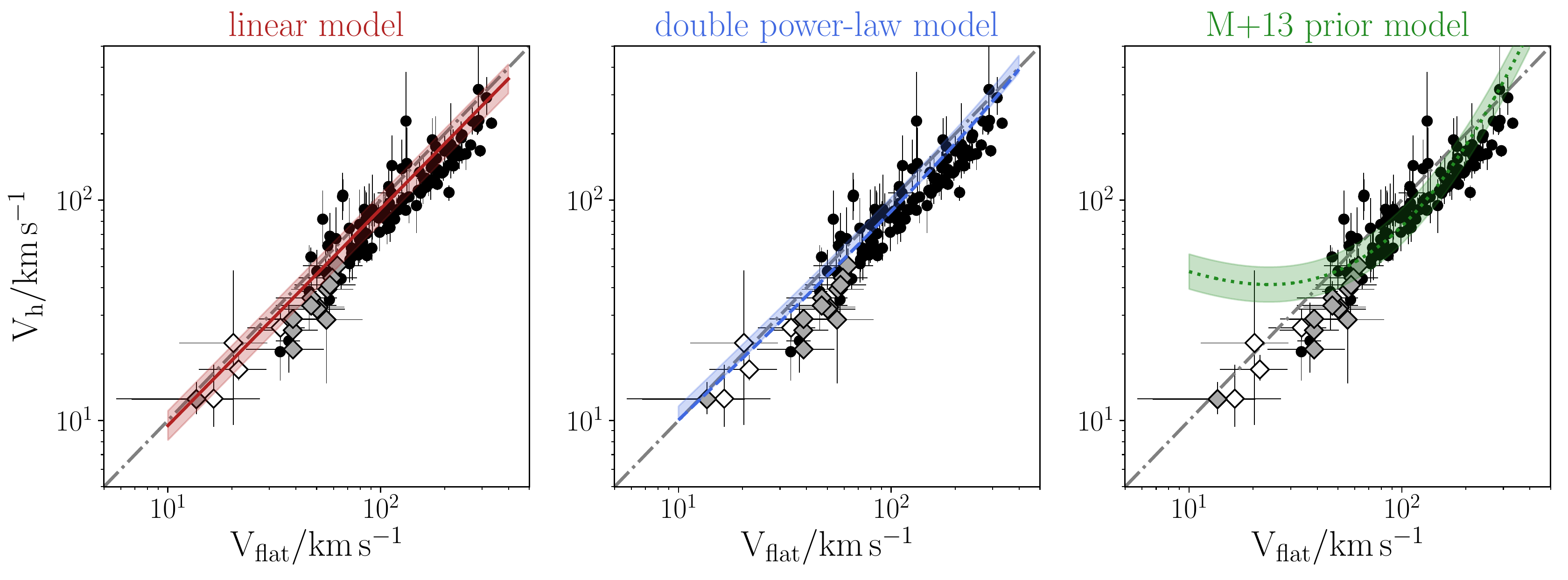}
\caption{Comparison of the predictions of the three models in the $\vf-\Vh$ plane, with
         data for the SPARC (circles) and LITTLE THINGS galaxies (diamonds). The
         halo virial velocities have been obtained with a careful rotation curve
         decomposition by \cite{Read+17} for the LITTLE THINGS galaxies and by
         \citetalias{PFM19} for the SPARC galaxies. In all panels, the grey dot-dashed
         line is the 1:1 and the symbols are the same as in Fig.~\ref{fig:fits}.
        }
\label{fig:comparison}
\end{figure*}

\subsection{Three fundamental fractions from dwarfs to massive spirals}

In Figure~\ref{fig:predictions} we show the predictions of the three best models
(on each column) of the three fundamental fractions, respectively $\fj$, $\fM$, and
$\fv$, as a function of $\vf$ (on each row).
The most important and most striking result to notice is that the predictions of
the three fractions behave similarly in the linear and double power-law models.
For the vast majority of galaxies
the predictions of these two models, which are by far statistically preferred to
the \citetalias{Moster+13} prior model, are in remarkable agreement, considering
that they have very different functional forms and degrees of freedom. The fact
that the agreement is so detailed in $\fM$, $\fj$, and $\fv$ ensures that the result
is robust and confirms that the data have enough information to infer these
fractions. This can, thus, be regarded as a major success of the modelling approach
presented in this work.

Along the same lines, another interesting result is that even when allowing the
behaviour of the three fractions to change slope at a characteristic velocity
($\lvfzero$), i.e. the parameters preferred by the data, $\fM$ and $\fv$ do not
have a significant break at the scale of Milky Way-sized galaxies. This is a
key prediction of abundance matching models. Considering that the best
\citetalias{Moster+13} prior model which breaks at $V_0\approx 125$ km/s is
statistically disfavoured, we conclude that the observed scaling laws of nearby
discs do not provide clear indications of any break in the behaviour of the
fundamental fractions at the scale of $L^\ast$ galaxies \citep[e.g.][]{McGaugh+10}.

Both the best linear and double power-law models have a global star
formation efficiency which grows monotonically with galaxy mass, approximately as
$\mstar^{1/3}$. Henceforth, the most
efficient galaxies at forming stars are the most massive spirals ($\mstar\gtrsim
10^{11}\Msun$, $\vf\gtrsim 250$ km/s), qualitatively confirming previous results
on detailed rotation curve decomposition \citepalias[][see Sec.~\ref{sec:rotcurve}
for a more in-depth comparison]{PFM19}. We also note that the most massive spirals
in both models have $\fM\sim\fb$, which implies that these systems have virtually no
missing baryons \citepalias{PFM19}.

The retained fraction of angular momentum is, on the other hand, remarkably
constant ($\fj\approx 0.6$) over the entire range probed by our galaxy sample
($\sim$1.5 dex in velocity, $\sim$5 dex in mass). Putting together our two
main findings on $\fM$ and $\fj$, we are now able to cast new light on why
disc galaxies today have comparable angular momenta to those of their dark haloes.
Since the slopes of the power-law $j-M$ relations for galaxies and haloes are
nearly the same within the uncertainties ($\sim 2/3$), then from
$\jstar\propto\fj\fM^{-2/3}\Mstar^{2/3}$ it follows that the factor
$\fj\fM^{-2/3}$ has to be nearly constant with mass \citep[e.g.][their Eq.s
15-16]{RF12}. This implies that the retained fraction of angular momentum
has to correlate with the global star formation efficiency
($\log\fj\propto\log\fM$) to reproduce the observed scalings.
Most of the earlier investigations on $\fj$ found that it was nearly
constant with mass \citep{DuttonvdBosch12,RF12,FR13,FR18}, since they
all adopted a monotonic $\fM$ \citep[from][]{Dutton+10}. \cite{Posti+18a},
instead,
used different models for the stellar-to-halo mass relation
to derive $\fj$ as a function of mass such that the observed Fall relation was
reproduced. Since most of the contemporary and popular models for
$\fM=\fM(\Mstar)$ have a bell shape, the constraint $\log\fj\propto\log\fM$
led these authors to conclude that a bell-shaped $\fj=\fj(\Mstar)$ was also favoured.
This, in turn, implies for instance that dwarfs should have significantly smaller
$\fj$ than $L^\ast$ galaxies \citep[e.g.][]{El-Badry+18,Marshall+19}.
However, the recent halo mass estimates by \citetalias{PFM19} indicated that
spirals are following a simpler stellar-to-halo mass relation, roughly
$\fM\propto\Mstar^{1/3}$. This, together with the constraint
$\fj\fM^{-2/3}\approx const.$ implies a very weak dependence of $\fj$ on
stellar mass, roughly $\fj\propto\Mstar^{2/9}$. The comprehensive analysis
presented in this paper confirms this and points towards an even weaker
dependence of $\fj$ on mass, which is consistent with this value being constant ($\fj\approx 0.6$)
within the scatter.

The velocity fraction $\fv$ is found to be always compatible with unity in
the linear and double power-law models. The expression
$\fv\approx 1$ means that discs are
rotating close to the halo virial velocity, which roughly matches what is
seen in hydrodynamical simulations at the high-mass end \citep[e.g.][]{Ferrero+17};
this is also supported by any reasonable mass
decomposition \citepalias[e.g.][]{PFM19}. Similar to the
case of $\fj$, $\fv$ also turns out to depend substantially on $\fM$.
If $\fM$ is monotonic then $\fv$ is also monotonic
and close to unity; while, if $\fM$ has a non-monotonic bell shape,
then $\fv$ also follows a similar behaviour, rapidly plunging below
unity for both dwarfs and high-mass spirals. Considering a Tully-Fisher
relation of the type $\vf\propto\Mstar^\delta$ and writing
$\vf\propto\fv\fM^{-1/3}\Mstar^{1/3}$ (see e.g. Appendix~\ref{app:mstar}),
then it follows that $\fv\propto\fM^{1/3}\Mstar^{\delta-1/3}$, which means
that roughly $\fv\propto\fM^{1/3}$ since the extra dependence on $\Mstar$
is very weak ($\delta\simeq 0.25-0.3$). The implication of this is that if
$\fM$ has a bell shape as expected from abundance matching models, then also
$\fv$ will have a similar shape \citep[e.g.][]{Cattaneo+14,Ferrero+17}. With our
comprehensive analysis we find that such models are statistically disfavoured
by the data, which instead favour a monotonically increasing $\fM$ and a
roughly constant $\fv\approx 1$.
We can now conclude that our results provide a simple and appealing explanation
to why the observed scaling laws are single, unbroken power laws: the
galaxy--halo connection is linear and the fractions \eqref{def:fracs} are
single-slope functions of velocity (or mass), instead of being complicated
non-monotonic functions which, when combined as in
Eqs.~\eqref{eq:stars_mv}-\eqref{eq:stars_jv},
conspire to yield power-law scaling relations.

To make sure that these results are not valid only for the SPARC+LITTLE
THINGS sample we considered, we repeated the same analysis on the
much larger galaxy sample from \cite{Courteau+07}. This sample contains
about 1300 spiral galaxies found in different environments and it has
a higher completeness than SPARC. However, the mass range is more limited
($8\lesssim\log\mstar/\Msun\lesssim 11.7$) and the data quality is
poorer, since it relies on optical (H$\alpha$) rotation curves, the
disc scale lengths are typically more uncertain and we have to use
estimator \eqref{def:jstarRF12} to compute $\jstar$ for all galaxies.
Nevertheless, when we built the three scaling relations and we fitted
the three models, we arrived at basically the same main conclusions
as above: the linear and double power-law models have very
similar predictions for $\fj$, $\fM$, and $\fv$ and they are statistically
preferred to the \citetalias{Moster+13} prior model. Thus, from this
test we conclude that the results we inferred on the fundamental fractions
using the SPARC+LITTLE THINGS sample are generally applicable for all
regularly rotating disc galaxies.

Finally, we note that assuming a linear or double power-law functional form
for the behaviour of the three fractions as a function of $\vf$ does not
bias our results. We tested this by fitting a non-parametric model,
where we do not assume any functional form for the behaviour of $\fj$, $\fM$,
and $\fv$ as a function of $\vf$. Instead, we bin the range in $\vf$ spanned
by the data with five bins of different sizes, such that the number of galaxies
per bin is roughly equal. We, thus, constrained the five values of $\fj$, $\fM$,
and $\fv$, together with the intrinsic scatter $\siglf$, for a grand total
of 16 degrees of freedom. The resulting predictions on the three fractions
are very well compatible with those of the linear or double power-law models;
we show these predictions in Appendix \ref{app:npmod}.

\subsection{Comparison of the predicted $\fM$ with detailed rotation curve
            decomposition}
\label{sec:rotcurve}

The three best models that we fitted to the stellar Tully-Fisher, size-mass,
and Fall relations, directly predict the virial masses of the dark matter haloes
hosting these spirals. Luckily all these galaxies have good enough photometric
and kinematic data to allow for an accurate decomposition of their observed
HI rotation curve, which can be used to get a robust measurement of their halo
masses. In particular, \citetalias{PFM19} and \cite{Read+17} have carefully
performed fits to the observed rotation curves for the SPARC and LITTLE
THINGS samples, respectively, and have provided measurements of $\Vh$.
We show in Figure~\ref{fig:fM-Mstar} the measurements of $\fM$ for
these galaxies. Since these measurements rely on fits of the dark matter halo
profile and since they have not been used in the model fit carried out in this
paper, we can now check a posteriori if the predictions of our three
best models agree with the global shape of the halo profile inferred from the
HI rotation curves.

We show this comparison in Figure~\ref{fig:comparison}, in which we plot the
observed $\vf$ against the $\Vh$ measured from the rotation curve decomposition;
predictions from the three best models are superimposed. The predictions
of the linear model are by far in best agreement with the measurements.

The double power law is in a similar remarkable agreement for all
galaxies. From this check we conclude that these two models both
provide a good description of the observed disc galaxy
population, but with a preference for the linear model from a
statistical point of view, i.e. from the standard statistical criteria AIC and BIC.
On the other hand, the \citetalias{Moster+13} prior model manifestly fails
at reproducing the measured distribution of galaxies in the $\vf-\Vh$ plane,
both at low masses and, possibly, at high masses. According to the predictions of this
model, both dwarfs and very massive spirals should inhabit much more massive
dark matter haloes than what it is suggested from their HI rotation curves.
This has already been noted and dubbed the ``too big to fail" problem in the
field \citep{Papastergis+15}.

Thus, we conclude that a simple empirical model (of the type
Eqs.~\ref{eq:stars_mv}-\ref{eq:stars_jv}), in which all disc galaxies follow
a stellar-to-halo mass relation which has a peak at $\mstar\sim 3\times
10^{10}\Msun$, predicts galaxy formation fundamental parameters that are
discrepant with measurements of the kinematics of cold gas in spirals.
A simple tight and linear galaxy--halo connection, in disagreement with
abundance matching, however fully cures this too big to fail problem.

\section{Conclusions} \label{sec:concl}

In this paper we used the observed stellar Tully-Fisher, size-mass, and Fall
relations of a sample of $\sim 150$ nearby disc galaxies, from dwarfs to massive
spirals, to empirically derive three fundamental parameters of galaxy
formation: the global star formation efficiency ($\fM$), the retained fraction
of angular momentum ($\fj$), and the ratio of the asymptotic rotation velocity
to the halo virial velocity ($\fv$).

Under the usual assumption that the galaxy size is related to its specific
angular momentum, we used an analytic model to predict the distribution of
discs in the mass-velocity, size-velocity, and angular momentum-velocity
planes. We defined three models with different parametrisations of how the
three fundamental parameters vary as a function of asymptotic velocity
(or galaxy mass): we thus tested a linear model, a double power-law model,
and another with a double power-law behaviour, but with prior imposed
such that the model follows the expectations from widely used abundance
matching stellar-to-halo mass relations for the global star formation
efficiency (the \citetalias{Moster+13} prior model).

We find the best-fitting parameters in each of these models and their
posterior probabilities performing a Bayesian analysis.
We briefly summarise our main findings:
\begin{itemize}
    \item We find reasonably good fits of the observed scaling relations in
          all three cases that we have tested.
    \item By assuming that the intrinsic scatter is the same for all three
          fundamental fractions (for computational simplicity), we find
          that this scatter has to be particularly small ($\siglf\simeq 0.07 \pm 0.01$
          dex) to account for the intrinsic scatters of the three
          observed scaling relations.
    \item We determined
          that the statistically preferred model is that where
          the fundamental galaxy formation parameters vary linearly with
          galaxy velocity (or mass) using standard statistical criteria (AIC \& BIC).
          On the other hand, the model with standard
          abundance matching priors \citepalias[from][]{Moster+13} is largely
          disfavoured by the data. We conclude that models where the galaxy--halo
          connection is complex and non-monotonic statistically provide
          an overfit to the structural scaling relations of discs.
    \item We empirically derive and show how the three fundamental parameters
          vary as a function of galaxy rotation velocity.  We find that in
          the best-fitting linear and double power-law models the
          three fractions have a remarkable similar behaviour, despite having
          completely different functional forms.
          This ensures that the observed scaling laws really provide a strong,
          data-driven inference on the galaxy--halo connection.
    \item We confirm previous indications that the retained fraction of angular
          momentum and the ratio of the asymptotic-to-virial velocity strongly
          depend on the global star formation efficiency
          \citep[e.g.][]{NavarroSteinmetz00,Cattaneo+14,Posti+18a}; in particular,
          they are non-monotonic only if the latter is non-monotonic.
    \item In the statistically preferred models, the retained fraction of
          angular momentum is relatively constant across the entire mass range
          ($\fj \sim 0.6$) as is the ratio of the asymptotic-to-virial
          velocity ($\fv \sim 1$). On the other hand, the global star formation
          efficiency is found to be a monotonically increasing function of
          mass, implying that the most efficient galaxies at forming stars
          are the most massive spirals (with $\fM \sim \fb$), whose star
          formation efficiency has not been quenched by strong feedback (the
          failed feedback problem).
    \item Finally, we compared a posteriori the predictions of the three
          models with the dark matter halo masses found by \cite{Read+17} and
          \citetalias{PFM19} from the detailed analysis of rotation curves in
          the LITTLE THINGS and SPARC galaxy samples. We found that the
          \citetalias{Moster+13} prior model is strongly rejected since it
          significantly overpredicts the halo masses especially at low $\vf$,
          but also at high $\vf$.  This too big to fail problem
          \citep{Papastergis+15} is fully solved in the linear model,
          which best describes the measurements.
\end{itemize}

Our analysis leads us to conclude that the statistically favoured
explanation to why the observed scaling laws of discs are single, unbroken
power laws is the simplest possible: the fundamental galaxy formation parameters
for spiral galaxies are tight single-slope monotonic functions of mass,
instead of being complicated non-monotonic functions.

The present study and the associated failed feedback problem concern only
disc galaxies. It is known that when including also spheroids, which dominate
the galaxy population at the high-mass end, the inferred galaxy-halo connection
becomes highly non-linear. In particular, it appears that there is a clear
difference in the stellar-to-halo mass relations for spirals and ellipticals
at least at the high-mass end, as probed statistically using many observables
\citep[e.g.][]{Conroy+07,Dutton+10,More+11,WojtakMamon13,Mandelbaum+16,Lapi+18}.
Thus, the results found in this work and those of \citetalias{PFM19} could in
principle be reconciled with conventional abundance matching if
the galaxy-halo connection is made dependent on galaxy type. This can be
achieved for instance if discs and spheroids have significantly different
formation pathways, i.e. in accretion history, environment etc., which
are still encoded today in their different structural properties
\citep[e.g. also][]{Tortora+19}. Whether this is the case in current
simulations of galaxy formation, and whether the failed feedback problem in massive discs can be addressed within those simulations is the next big
question to be asked.

The model preferred by the SPARC and LITTLE THINGS data has a monotonic $\fM$
approximately proportional to $\mstar^{1/3}$.
With this global star formation efficiency, it turns out that the retained
fraction of angular momentum $\fj$ needs to be high and relatively constant
for discs of all mass ($\fj\approx 0.5-0.9$). This is because the observed
Fall relation has a similar slope to the specific angular momentum-mass
relation of dark matter haloes: since $\fj\approx const$, a simple
correspondence $\jstar\propto\jh$ is in remarkable agreement with the
observations \citep[][]{RF12,Posti+18a}. This implies that the current
measurements are compatible with a model in which discs have overall retained
about all the angular momentum that they gained initially from tidal torques.
This is to be intended in an integrated sense in the entire galaxy: it can not
simply have happened with every gas element having conserved their angular
momentum (sometimes referred to as ``detailed'' or ``strong'' angular momentum
conservation) because dark matter haloes and discs have completely different
angular momentum distributions today \citep[][]{Bullock+01,vandenBosch+01}.
Thus, even if stars and dark matter appear to have a simple correspondence
$\jstar\propto\jh$, it remains unclear and unexplained how the angular
momentum of gas and stars has redistributed during galaxy formation and why
the total galaxy's specific angular momentum is still proportional to that of
its halo.

Some disc galaxies, especially dwarfs, have huge cold gas reservoirs which
sometimes dominate over their stellar budget. These systems
are typically outliers of the (stellar) Tully-Fisher, but they instead lie
on the baryonic Tully-Fisher relation, which is obtained by replacing the
stellar mass with the baryonic mass \citep[$\mbary=\mstar+\mhi$, see e.g.][]
{McGaugh+00,Verheijen01}. More galaxies adhere to this relation,
which is also tighter than the stellar Tully-Fisher, suggesting that it
is a more fundamental law \citep[e.g.][]{Lelli+16a,Ponomareva+18}. Thus,
considering baryonic fractions instead of stellar fractions (Eq.
\ref{def:fracs}) in a model such as that of Sec.~\ref{sec:model} would
presumably give us deeper and more fundamental insight into how baryons
cooled and formed galaxies. To do this, it is thus imperative to have a
baryonic counterpart of the size-mass and Fall relations \citep[e.g.][]
{ObreschkowGlazebrook14,Kurapati+18}, for a sample of spirals sufficiently
large in mass. We plan to report on the latter soon, establishing first
whether the observed baryonic Fall relation is tighter and more fundamental
than the stellar Fall relation.

\begin{acknowledgements}
We thank the anonymous referee for an especially careful and constructive
report. LP acknowledges support from the Centre National d’Etudes Spatiales
(CNES). BF acknowledges support from the ANR project ANR-18-CE31-0006.

\end{acknowledgements}

\bibliographystyle{aa} 
\bibliography{refs} 

\newpage
\appendix
\section{Size and angular momentum fractions in disc galaxies} \label{app:fr}

The specific angular momentum of an exponential disc with a flat rotation curve
$\vf=\fv\Vh$ is
\begin{equation} \label{eq:app1}
    \jstar = \frac{\int\de R\, R^2\,\exp\left(-R/\rd\right)\,\vf}{\int\de R\, R\,\exp\left(-R/\rd\right)}
           = 2\rd\fv\Vh.
\end{equation}
From Eqs.~\eqref{eq:dm_rv}-\eqref{eq:dm_jv} and introducing $\fj$ and $\fr$ as
in Eq.~\eqref{def:fracs}, we have
\begin{equation} \label{eq:app2}
    \rd = \frac{\fr}{H}\sqrt{\frac{2}{\Delta}} \Vh;
\end{equation}
\begin{equation} \label{eq:app3}
    \jstar = \frac{2\lambda\fj}{H\sqrt{\Delta}} \Vh^2.
\end{equation}
Plugging Eq.~\ref{eq:app1} into Eq.~\ref{eq:app3} we obtain
\begin{equation} \label{eq:app4}
    \rd = \frac{\fj\lambda}{\fv H\sqrt{\Delta}} \Vh,
\end{equation}
which, using Eq.~\ref{eq:app2}, can be rearranged as
\begin{equation} \label{eq:app5}
    \fr = \frac{\lambda}{\sqrt{2}} \frac{\fj}{\fv}.
\end{equation}
A very similar relation to this was already derived by
\cite{FallEfstathiou80} and \cite{MMW98}, who started by assuming that
$\jstar=2\rd\Vh$ to replace Eq.~\eqref{eq:app1} and got $\fr =
\lambda\fj/\sqrt{2}$. In this work we are interested in deriving simultaneous
constraints on $\fj$,$\fM$, and $\fv$, thus we prefer to use the formulation
in Eq.~\eqref{eq:app5}, which allows the flat asymptotic circular velocity $\vf$
to differ from the halo virial velocity $\Vh$. We have, however, checked that
using $\fr = \lambda\fj/\sqrt{2}$ instead of Eq.~\eqref{eq:app5} does not
significantly alter the fits of the models.

\section{Fitting $\fj$, $\fM$, and $\fv$ as a function of $\mstar$} \label{app:mstar}

In this Appendix we demonstrate that considering $\mstar$ as the independent observable,
and thus fitting the canonical Tully-Fisher, size-mass, and Fall relations, yields similar
predictions for the fractions $\fj$, $\fM$, and $\fv$ to what we obtained above.

We start from the equations for dark matter, i.e.
\begin{equation} \label{eq:APPdm_vm}
    \Vh = \left(\sqrt{\frac{\Delta}{2}} GH\Mh\right)^{1/3};
\end{equation}
\begin{equation} \label{eq:APPdm_rm}
    \Rh = \left( \frac{2G\Mh}{\Delta H^2} \right)^{1/3};
\end{equation}
\begin{equation} \label{eq:APPdm_jm}
    \jh = \frac{\lambda}{(\Delta H^2)^{1/6}}\left( 2G\Mh\right)^{2/3}.
\end{equation}

After introducing the three fractions, we have
\begin{equation} \label{eq:APPstars_vm}
    \vf = \fv\left(\sqrt{\frac{\Delta}{2}} \frac{GH\mstar}{\fM}\right)^{1/3};
\end{equation}
\begin{equation} \label{eq:APPstars_rm}
    \rd = \frac{\lambda\fj}{\fv}\left( \frac{G\mstar}{\sqrt{2}\Delta H^2\fM} \right)^{1/3};
\end{equation}
\begin{equation} \label{eq:APPstars_jm}
    \jstar = \frac{\lambda\fj}{(\Delta H^2)^{1/6}}\left( \frac{2G\mstar}{\fM}\right)^{2/3}.
\end{equation}
The three equations above are used to fit the observations in the $\vf-\mstar$,
$\rd-\mstar$ and $\jstar-\mstar$ diagrams.

Similar to Sec.~\ref{sec:fractions}, we try three different models as follows:\begin{itemize}
    \item[{\bf (i)}] the linear model, where
                     \begin{equation} \label{eq:APPlin}
                        \log\,f = \alpha\log\Mstar/\Msun + \lfzero.
                     \end{equation}
    \item[{\bf (ii)}] the double power-law model, where
                     \begin{equation} \label{eq:APPdpl}
                        f = \fzero \left(\frac{\mstar}{\mzero}\right)^\alpha
                        \left(1+\frac{\mstar}{\mzero}\right)^{\beta-\alpha}.
                     \end{equation}
    \item[{\bf (iii)}] the \citetalias{Moster+13} prior'' model, which is the same as
                       case (ii), but with priors on $\fM$ from the abundance
                       matching model of \citetalias{Moster+13}.
\end{itemize}

\begin{figure*}
\includegraphics[width=\textwidth]{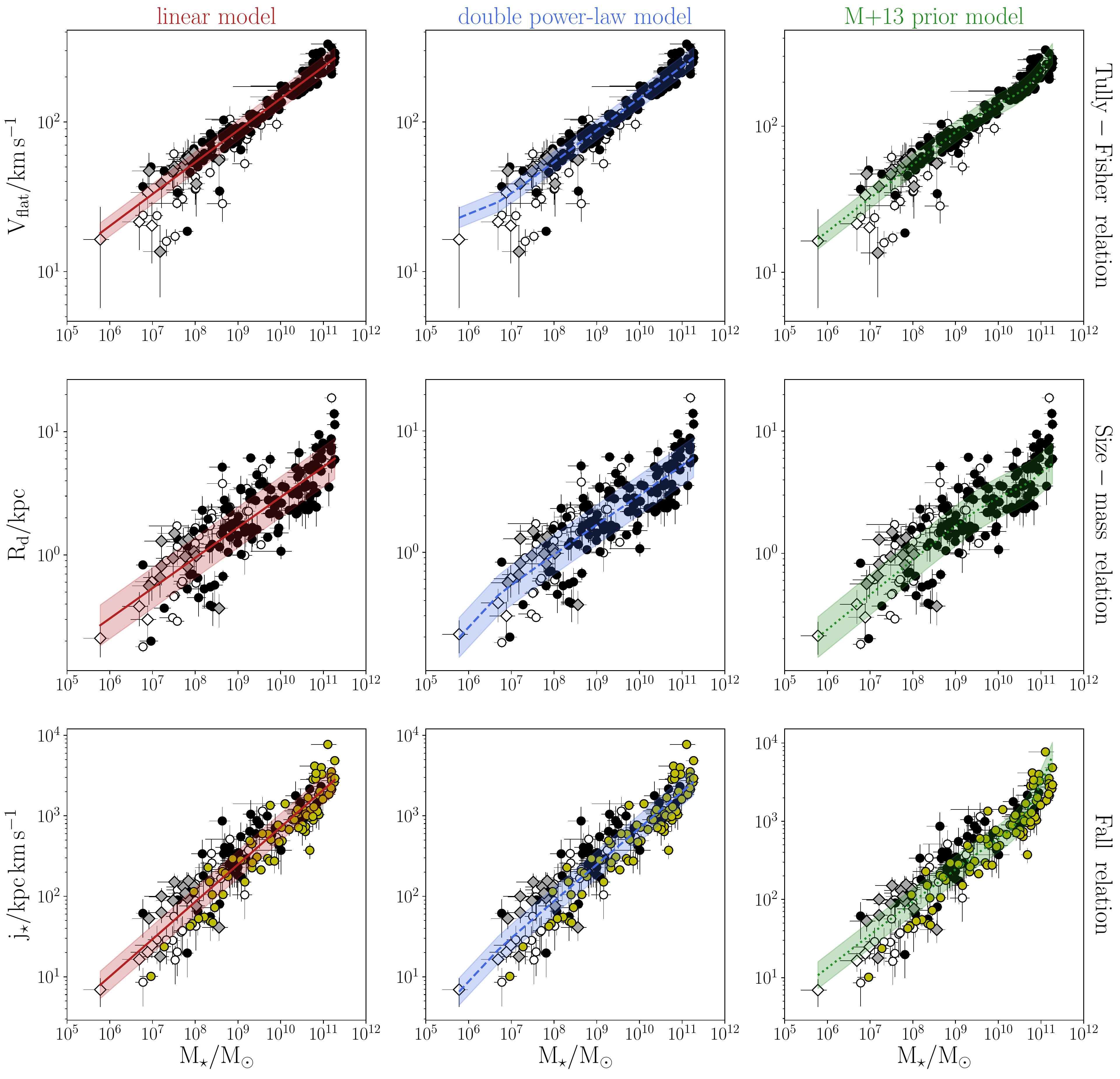}
\caption{Comparison of the three models described in Appendix~\ref{app:mstar} with the
         data on the Tully-Fisher, size-mass, and Fall diagrams. Curves and symbols
         are analogous to Fig.~\ref{fig:fits}.
        }
\label{fig:mstar_fits}
\end{figure*}

\begin{figure*}
\includegraphics[width=\textwidth]{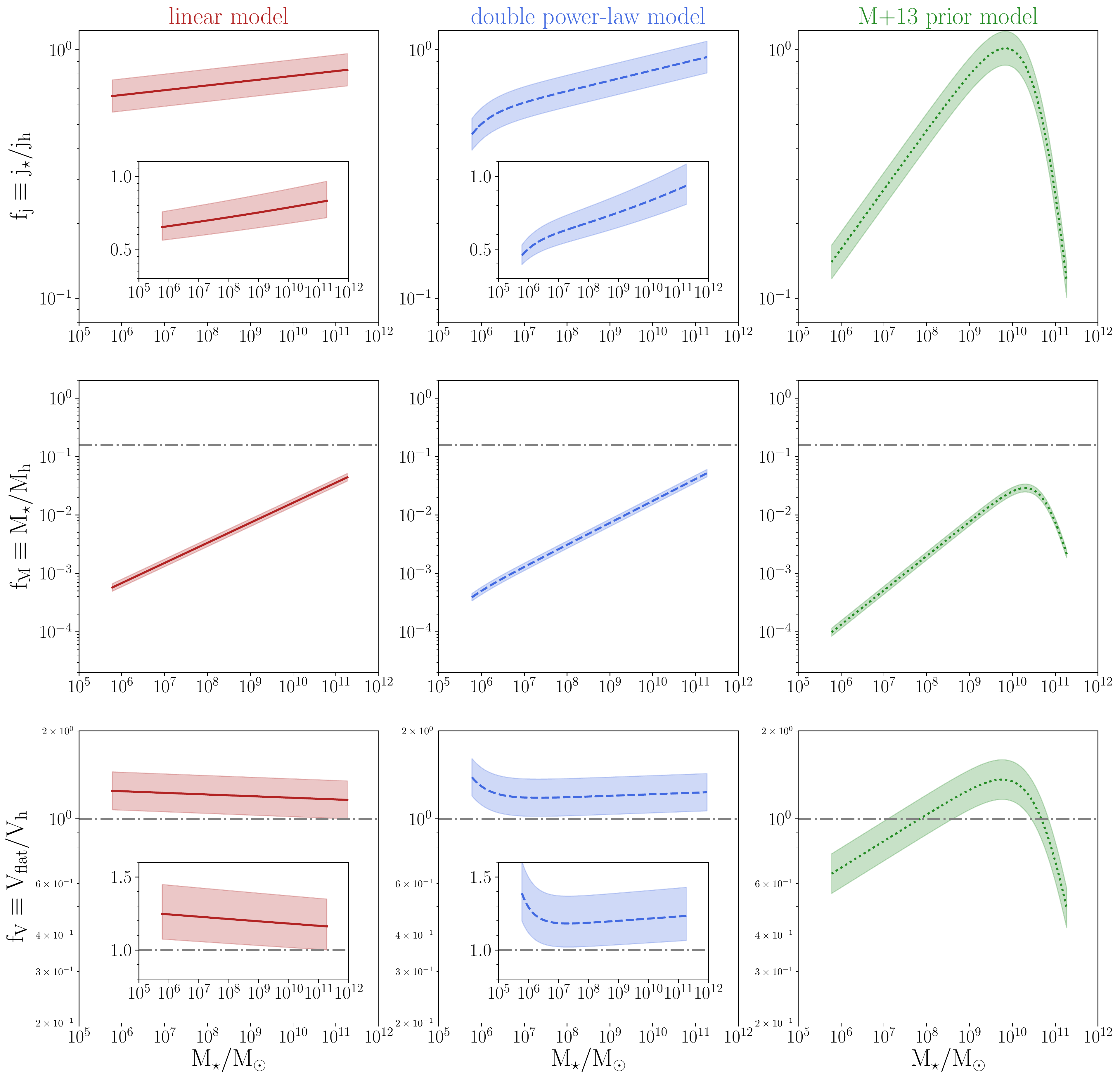}
\caption{Behaviour of the three fractions as a function of $\mstar$ in the three models
         of Appendix~\ref{app:mstar}. Curves and symbols are analogous to
         Fig.~\ref{fig:predictions}.}
\label{fig:mstar_predictions}
\end{figure*}

We show in Figure~\ref{fig:mstar_fits}-\ref{fig:mstar_predictions}, which are analogous
to Fig.~\ref{fig:fits}-\ref{fig:predictions}, the data/model comparisons and the
constraints on the fundamental fractions, respectively. The behaviour of the models
is largely identical to those in the main text; the only noticeable difference
is that the double power-law model now has a slight break at low masses
(at $\sim 10^6\Msun$). However the statistical significance of this break is very low
since it is driven by just a few data points in the dwarf regime, where the
uncertainties are higher. On the other hand, for this model we still find no indications
of a break at around $L^\ast$ galaxies.
Finally, we note that also in this case we find that the statistically preferred model
is the linear model, according to the AIC \& BIC criteria: $(\Delta{\rm AIC},
\Delta{\rm BIC}) = (1.4, 13.3)$ with respect to the double power-law model and
$(\Delta{\rm AIC}, \Delta{\rm BIC}) = (28.2, 40.1)$ with respect to the \citetalias{Moster+13} prior model.

\begin{table}
\caption{Posterior distributions for the three models with $\mstar$ as the main
independent observable quantity. This Table is analogous to Tab.~\ref{tab:parameters}
and the models and the equations fitted described in App.~\ref{app:mstar}.}
\label{tab:APPparameters}
\begin{center}
\setlength\extrarowheight{5pt}
\begin{tabular}{@{\extracolsep{\fill}} lr|r|r}
& linear & double power law & \citetalias{Moster+13} prior\\
\hline\hline
 $\log\,f_{0,j}$ & $-0.36^{+0.57}_{-0.57}$ & $-0.33^{+0.34}_{-0.37}$ & $0.33^{+0.07}_{-0.06}$ \\[5pt]
 $\log\,\mzero/\Msun$ & -- & $3.6^{+0.6}_{-0.4}$ & $10.76^{+0.04}_{-0.05}$ \\[5pt]
 $\alpha_{j}$ & $0.02^{+0.06}_{-0.06}$ & $2^{+25}_{-26}$ & $0.25^{+0.04}_{-0.04}$ \\[5pt]
 $\beta_{j}$ & -- & $0.04^{+0.06}_{-0.06}$ & $-1.9^{+0.15}_{-0.15}$ \\[5pt]
\hline
 $\log\,f_{0,M}$ & $-5.3^{+0.7}_{-0.7}$ & $-4.1^{+0.4}_{-0.5}$ & $-1.1^{+0.05}_{-0.05}$ \\[5pt]
 $\alpha_{M}$ & $0.35^{+0.07}_{-0.07}$ & $3^{+25}_{-27}$ & $0.61^{+0.04}_{-0.04}$ \\[5pt]
 $\beta_{M}$ & -- &$0.4^{+0.07}_{-0.07}$ & $1.63^{+0.05}_{-0.05}$ \\[5pt]
\hline
 $\log\,f_{0,V}$ & $0.11^{+0.24}_{-0.24}$  & $0.06^{+0.14}_{-0.15}$  & $0.26^{+0.03}_{-0.03}$ \\[5pt]
 $\alpha_{V}$ & $0.01^{+0.02}_{-0.02}$ & $-11^{+25}_{-19}$ & $0.1^{+0.02}_{-0.02}$ \\[5pt]
 $\beta_{V}$ & -- & $0.01^{+0.02}_{-0.02}$ & $-0.87^{+0.07}_{-0.07}$ \\[5pt]
\hline
 $\sigma_{\log\,f}$ & $0.07^{+0.01}_{-0.01}$ & $0.07^{+0.01}_{-0.01}$ & $0.08^{+0.01}_{-0.01}$\\[5pt]
\end{tabular}
\end{center}
\end{table}

\section{Non-parametric model} \label{app:npmod}
In this Appendix, we describe a model in which the variation of $\fj$, $\fM$, and $\fv$ as a function
of $\vf$ has a completely free form. This model is aimed to test whether the functional
forms that we have chosen in Sec.~\ref{sec:fractions} are too restrictive for the data
that we considered, and if the data themselves are informative enough to constrain
a different behaviour.

We binned the range in $\vf$ spanned by the data, [15,320]
km/s, into five bins of different sizes, such that the number of galaxies in each bin is
roughly equal. We, then, constrained the five (constant) values of $\fj$, $\fM$, and $\fv$
in each bin maximising the same likelihood as in Sec.~\ref{sec:like}. Together with
the intrinsic scatter $\siglf$, this model has a total of 16 degrees of freedom.

\begin{figure*}
\includegraphics[width=\textwidth]{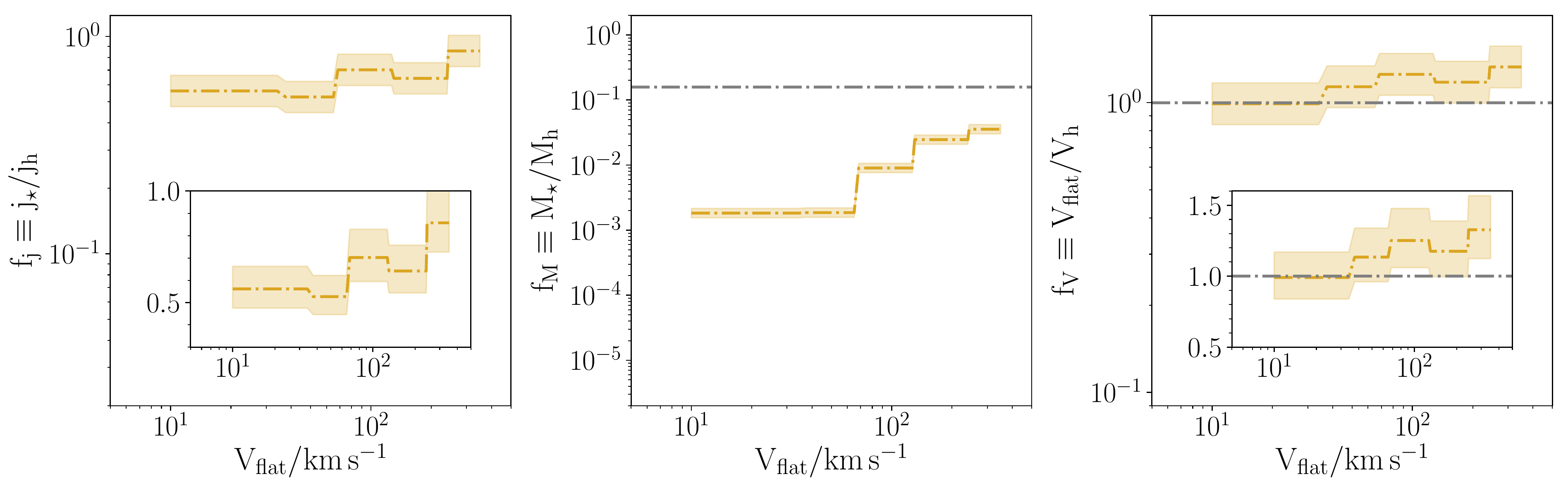}
\caption{Fundamental fractions as a function of the asymptotic rotation velocity for the
         non-parametric model (gold dot-dashed lines, with the band encompassing the
         intrinsic scatter $\siglf$). The three fractions are binned in 5 bins in
         $\vf$ (with roughly the same number of galaxies). The panels are in the same
         scale as those in Fig.~\ref{fig:predictions}.
        }
\label{fig:fractions_NPMod}
\end{figure*}

Figure~\ref{fig:fractions_NPMod} shows the three fractions in this model as a function
of $\vf$. A part from a small difference in the lowest $\vf$ bin, where dwarf galaxies
with the highest uncertainties dominate, the predictions of this model are in remarkable
agreement with those of the linear and double power-law models. Also the
intrinsic scatter that we fit with this model is very well comparable with that
of the other cases, where $\siglf=0.08 \pm 0.05$. This ensures that
the linear or double power-law functional forms that we adopted for our
fiducial models are not too restrictive for the data that we have at hand.

\end{document}